\documentclass[onecolumn,authoryear]{els-mrw} 

\usepackage{amsmath,amssymb,amsfonts,amsthm,makeidx,graphicx}
\usepackage{txfonts}
\usepackage{helvet}
\usepackage{xspace}
\usepackage[acronym,shortcuts]{glossaries}
\usepackage[dvipsnames]{xcolor} 
\usepackage[colorlinks=true,linkcolor=blue,urlcolor=blue,citecolor=blue]{hyperref}
\usepackage{siunitx}
\usepackage{ulem}

\graphicspath{{./}, {figures/}}

\newcommand{\rsun}{\ensuremath{\,\rm{R}_{\odot}}\xspace}
\newcommand{\km}{\ensuremath{\,\rm{km}}\xspace}

\newcommand{\msun}{\ensuremath{\,\rm{M}_{\odot}}\xspace}

\newcommand{\lsun}{\ensuremath{\,\rm{L}_{\odot}}\xspace}
\newcommand{\eg}{e.g.\@\xspace}
\newcommand{\cf}{c.f.\@\xspace}
\newcommand{\ie}{i.e.\@\xspace}
\renewcommand{\emph}[1]{\textit{#1}}
\DeclareUnicodeCharacter{3B1}{\ensuremath{\alpha}}
\DeclareUnicodeCharacter{3B7}{\ensuremath{\eta}}
\newacronym{CE}{CE}{common envelope}
\newacronym{RLOF}{RLOF}{Roche-lobe overflow}
\newacronym{MS}{MS}{main-sequence}
\newacronym[longplural={luminous red novae}, shortplural={LRNe}]{LRN}{LRN}{luminous red nova}

\newcommand{\chapterreft}[1]{see also the Chapter on ``#1'' in this Encyclopedia\xspace}
\newcommand{\chapterref}[1]{(\chapterreft{#1})\xspace}

\begin{document}

\chapter{Stellar mergers and common-envelope evolution}\label{chap1}

\author[1,2,4]{Fabian~R.\ N.\ Schneider}%
\author[1,5]{Mike Y.\ M.\ Lau}%
\author[1,2,3,6]{Friedrich K.\ R{\"o}pke}%
\address[1]{\orgname{Heidelberger Institut f{\"u}r Theoretische Studien}, \orgaddress{Schloss-Wolfsbrunnenweg 35, 69118 Heidelberg, Germany}}
\address[2]{\orgname{Zentrum f{\"u}r Astronomie der Universit{\"a}t Heidelberg}, \orgdiv{Astronomisches Rechen-Institut}, \orgaddress{M{\"o}nchhofstr.\ 12-14, 69120 Heidelberg, Germany}}
\address[3]{\orgname{Zentrum f{\"u}r Astronomie der Universit{\"a}t Heidelberg}, \orgdiv{Institut f{\"u}r Theoretische Astrophysik}, \orgaddress{Philosophenweg 12, 69120 Heidelberg, Germany}}
\address[4]{{\tt fabian.schneider@h-its.org}}
\address[5]{{\tt mike.lau@h-its.org}}
\address[6]{{\tt friedrich.roepke@h-its.org}}
%

\maketitle

\begin{glossaryA}[Glossary]
    \term{CNO cycle:} Cycle of fusion reactions during which hydrogen is converted into helium using carbon (C), nitrogen (N) and oxygen (O) as catalysts. This burning mode is the main nuclear energy provider in main-sequence stars initially more massive than ${\approx}\,1.3\,\msun$. During CN(O) cycling, the nitrogen abundance is typically increased at the expense of carbon and oxygen nuclei. 

    \term{Convection:} Process in which energy/heat is transported by the motion of large fluid parcels. In stars, this type of energy transport takes over whenever radiative photon diffusion (radiative energy transport) is insufficient.

    \term{Dynamical timescale:} Free-fall or sound-crossing timescale of a star. The dynamical timescale is related to the binary star's orbital period in the cases of stellar mergers and common-envelope phases. For the Sun, this timescale is about half an hour. This is the shortest of the three main stellar timescales (\cf thermal and nuclear timescales).

    \term{Eddington limit:} This limit describes a force balance between outward photon pressure and inward (self) gravitational pressure in spherical symmetry in, \eg, a star. Often, electron scattering opacity is assumed to compute the photon pressure but the opacity can be more complex.

    \term{Stellar envelope:} Stars are often separated into ``core'' and ``envelope''. These terms, however, are sometimes not clearly defined. In this article, we refer to the ``envelope'' as the hydrogen-rich outermost layers of stars. Further in, there are then layers enriched in various chemical elements produced by previous nuclear burning (\eg helium, carbon and oxygen).

    \term{Lagrange points:} Saddle points and local maxima in the effective Roche potential, \ie locations at which a test particle would experience no net forces. The Roche potential is defined for two point particles (the binary stars) in a co-rotating frame and includes the gravitational and centrifugal potentials. 

    \term{Magnetic dynamo:} A process in which a seed magnetic field is rapidly amplified by fluid motions and other (magnetohydrodynamic) instabilities. 

    \term{Main sequence (star):} Phase in the evolution of stars where they undergo nuclear fusion of hydrogen into helium in their cores. This phase amounts to about 90\% of a stellar lifetime.

    \term{Mean molecular weight:} The dimensionless mean molecular weight $\mu$ is the mean mass per particle in a gas in terms of the atomic mass unit $m_\mathrm{u}$. Thus, the average number of particles per unit volume in a gas is related to the mass density $\rho$ via $\left< n \right> = \rho/\mu m_\mathrm{u}$. In a fully ionized gas, particles include the atomic nuclei and electrons.

    \term{Nuclear timescale:} Timescale on which nuclear fuel is consumed. Most often, this relates to core hydrogen burning, \ie the main-sequence lifetime in stellar evolution. For the Sun, this timescale is about 10 billion years. This is the longest of the three main stellar timescales (\cf dynamical and thermal timescale). 

    \term{Planetary nebula:} Illuminated outer layers of low-to-intermediate mass stars ($0.8\text{--}8\,\msun$) that were ejected towards the end of their lives. The hot, leftover core of the star ionises the nebula. Pictures of planetary nebulae belong to the most iconic photographs in astronomy and have inspired many people. There are various mechanisms that can eject matter from stars and common-envelope evolution is one of them.

    \term{Post-main-sequence (star):} Phase in stellar evolution after exhausting the hydrogen fuel in the centre of stars. This phase includes shell hydrogen burning, core helium burning, and all other advanced nuclear burning phases.

    \term{pp-chain:} Nuclear reaction chains that effectively convert four hydrogen nuclei into one helium nucleus. The pp-chains dominate nuclear energy production in stars initially less massive than ${\approx}\,1.3\,\msun$.

    \term{Thermal timescale:} Timescale on which the entire internal energy of a star is radiated away or, in other words, the time it takes a star to reach full thermal equilibrium, \ie a state in which the surface energy losses are balanced by internal energy generation. In the Sun, this timescale is about 15 million years, which is longer than the dynamical timescale but shorter than the nuclear timescale. 
\end{glossaryA}

\begin{glossaryA}[Nomenclature]
    \begin{tabular}{@{}lp{34pc}@{}}
    CE & Common envelope \\
    MS & Main sequence\\
    LRN(e) & Luminous red nova(e) \\
    RSG & Red supergiant\\
    \end{tabular}
\end{glossaryA}

\begin{abstract}[Abstract]
Stellar mergers and common-envelope evolution are fast (dynamical-timescale) interactions in binary stars that drastically alter their evolution. They are key to understanding a plethora of astrophysical phenomena. Stellar mergers are thought to produce blue straggler stars, blue supergiants, and stars with peculiar rotation and surface chemical abundances. Common-envelope evolution is proposed as a key stage in the formation of gravitational wave sources, X-ray binaries, type Ia supernovae, cataclysmic variables, and other systems. A significant fraction (tens of percent) of binary stars undergo such a phase during their evolution. In this chapter, we first discuss processes leading to a stellar merger or common-envelope phase. We then explain these complex interactions, starting from underlying physical principles like entropy sorting in stellar mergers and the energy formalism in common envelopes. This is followed by a more complete picture revealed by three-dimensional (magneto)hydrodynamical simulations. The outcomes of these interactions are discussed comprehensively and special emphasis is given to the role of magnetic fields. Both stellar mergers and common-envelope evolution remain far from fully understood, and we conclude by highlighting open questions in their study.
\end{abstract}

\textbf{Keywords:} Common-envelope evolution -- Blue straggler stars  -- Magnetic fields -- Magnetic stars -- Magnetohydrodynamical simulations -- Multiple star evolution -- Stellar mergers -- Stellar outflows

\begin{BoxTypeA}[chap1:keypoints]{Key points}
        \begin{itemize}
            \item Stellar mergers and \ac{CE} evolution are processes that drastically alter a binary star on the short, dynamical timescale. Stellar mergers combine two stars into one, while \ac{CE} evolution has the potential to shrink a binary orbit by orders of magnitude while removing the envelope of a giant star.
            \item Pathways leading to a stellar merger or a \ac{CE} interaction include dynamically stable and unstable mass transfer, the Darwin instability, secular interactions in triples, and dynamical interactions in dense stellar environments.
            \item Historically, stellar mergers were first discussed in the context of collisions in dense stellar environments. They can give rise to blue and red stragglers, highly magnetic stars, and long-lived blue supergiants; many of the most massive stars in the Universe may in fact be merger products. At the end of their evolution, merged stars may explode in unusual supernovae such as SN~1987A and interacting supernovae, and they can collapse into very massive black holes. 
            \item The conception of \ac{CE} evolution was motivated by the separation problem, the observation that certain compact binaries have orbital periods that are too small to accommodate the sizes of their giant progenitors. \ac{CE} evolution is believed to be a key step in forming cataclysmic variables, gravitational wave sources, X-ray binaries, some planetary nebulae, and many more astrophysical objects.
            \item Stellar mergers and \ac{CE} evolution have been modelled using different methods with varying complexity, ranging from entropy sorting (for stellar mergers) and the energy formalism (for \ac{CE} evolution) to 1D and 3D (magneto-)hydrodynamic simulations. However, our understanding of the details of the processes is still limited and the wide range of spatial and temporal scales challenges numerical simulations.
            \item Stellar mergers and \ac{CE} interaction have been proposed to give rise to a class of transients broadly known as luminous red novae. 
        \end{itemize}
\end{BoxTypeA}

\section{Introduction}\label{chap1:introduction}

The lives and deaths of stars are influenced by a multitude of physical processes occurring on a wide range of timescales. Most stars do not change significantly over observable timescales because their evolution is dominated by long nuclear burning processes. For the Sun, this almost static phase lasts for billions of years. In binary systems or higher-order multiples, stellar mergers and \ac{CE} evolution are processes that could give rise to drastic changes on the much shorter dynamical timescale, which is of the order of the orbital period (lasting from days to years). In a stellar merger, two stars combine to form a single object. In \ac{CE} evolution, two stellar objects orbit each other inside a low-density envelope that originally belonged to one of them\footnote{A special case arises when two giant stars merge and the common envelope forms from the material of both initial envelopes, sometimes called a double-core \ac{CE}.}. Relative motion between the objects and this envelope induces a drag force that shrinks the orbital separation by orders of magnitude. In both processes, the physical interaction is, to first order, driven by hydrodynamic and gravitational forces, which operate on the dynamical timescale. Other relevant physics including radiation transport, magnetohydrodynamics, and nuclear reactions may also play an important role and are being actively investigated. The short-lived nature of stellar mergers and \ac{CE} evolution makes their direct observation rather rare. As a result, much of our understanding originates from theoretical models and observations of objects believed to have experienced these processes.

Stellar mergers and \acp{CE} are not always clearly distinguished concepts. The main ambiguity arises from the fact that \ac{CE} evolution could end with the two stellar objects merging inside the envelope. This outcome is sometimes even referred to as a ``failed'' \ac{CE} or a \ac{CE} merger. Stellar mergers and \ac{CE} evolution are therefore loosely distinguished by the \emph{potential} for the latter to preserve binarity at the end of the dynamical interaction. This expectation stems from one of the stars in the \ac{CE} possessing a clearly distinguished core and envelope. The tenuous gaseous envelope may be ejected during the spiral-in, removing the source of drag and allowing the stellar cores to resettle into a stable, albeit much tighter, orbit. Stellar mergers generally refer to a coalescence of stars without clearly distinguished or loosely bound envelopes, such that merging into one object could not be averted. Ultimately, the degree of core-envelope separation varies continuously with stellar properties such as age, and there are ambiguous situations in which the distinction between mergers and \acp{CE} is not meaningful.

Finally, we note that the term ``common envelope'' is also frequently applied to (over)contact binaries. In these binaries, both stars are so large that their surfaces are in direct contact. Yet, they can avoid a dynamical interaction and remain stable over much longer timescales\footnote{For example, so-called W~UMa stars are low-mass contact binaries for a good fraction of their main-sequence lifetime.}. They are not the focus of this chapter. The scope of this chapter also excludes mergers between two compact objects (black holes, neutron stars, and white dwarfs), which are explored separately \chapterref{Observed Gravitational-Wave Populations}.

\section{Evolution leading to dynamical interaction}\label{chap1:contact-binaries}

The key process preceding and initialising a dynamical interaction, whether it be a merger or a \ac{CE} event, is the \textit{loss of co-rotation} in a binary system where both stars have been brought to physical contact. This means that the rotational frequency of at least one star differs from the orbital frequency, which induces drag forces against the orbital motion that can lead to a spiral-in. There are also binary systems that never established co-rotation prior to the onset of dynamical interaction. Similarly, material brought outside the equipotential of the outer Lagrange point cannot be in co-rotation with the binary and will lead to drag forces that can induce orbital shrinkage. In the following, we introduce several pathways that can lead to contact binaries and possibly to stellar mergers and \ac{CE} events \citep[for more details, see \eg][]{henneco2024a}.
\\ \\
\noindent\textbf{Stable mass transfer}
Stars grow in size during most phases of their evolution. In a binary star, this growth is limited by each star's \emph{Roche lobe}, which is the teardrop-shaped region within which material is gravitationally bound to one star \chapterref{Binary stars}. Stellar material that exceeds the Roche lobe is no longer confined to its star. This may initiate \emph{mass transfer}, where material flows from one star, called the \emph{donor}, to the other, called the \emph{accretor}. Generally, this process may not be fully conservative, meaning that some mass is lost from the binary altogether. Mass transfer can lead to contact in essentially two different ways. Firstly, as the accretor gains mass, it also grows in size accordingly, which could form a contact binary. Such a contact binary could be initially stable, but may lead to a dynamical merger once both stars continue to expand in their normal evolution. Additionally, the accretor may inflate a few to a few hundred times, much larger than its equilibrium size, if it cannot radiate away the added gravitational energy from accretion fast enough \citep{kippenhahn1977a, lau2024a}. Secondly, accretion may be severely limited if the added material carries too much angular momentum. A \ac{MS} star reaches its breakup rotation rate after accreting less than a few percent of its mass with the specific angular momentum of the Keplerian orbit at its surface \citep{packet1981a}. Additional material cannot be added at rest relative to the surface. The non-accreted mass might then cause a drag force or be lost through the outer Lagrange point, removing angular momentum from the binary system and possibly leading to contact. 
\\ \\
\noindent\textbf{Dynamically unstable mass transfer} 
Under certain conditions, mass transfer is \emph{dynamically unstable}. This means that the mass-loss rate increases with the amount of mass that is lost on the dynamical timescale, leading to runaway. This tends to happen for donor stars that are significantly evolved such that they possess deep convective envelopes that expand in response to mass loss. Dynamical instability is also more likely the greater the donor star's mass relative to the companion since mass transfer from a more massive to less massive object shrinks the orbital separation. Some material escapes the binary system as the companion star cannot accommodate an arbitrarily large mass-transfer rate. The loss of orbital angular momentum through this material brings the two stars into contact.
\\ \\
\noindent\textbf{Darwin instability}
Tidal instabilities can also cause a binary to come into contact. For a single star, its rotation slows as it expands according to angular momentum conservation. However, in a binary system, tidal forces compete with this process, and can instead spin up a star using angular momentum extracted from the orbit until the orbit and the stellar spins are synchronised \chapterref{Tides}. However, a smaller orbit also has a larger orbital frequency. It is then possible that a giant star becomes sufficiently large that its moment of inertia prevents it from ever spinning up enough to catch up with the simultaneously rising orbital frequency. It has been shown that this occurs when the star's spin angular momentum is greater than a third of the orbital angular momentum \citep{hut1980a}. In this scenario, tides shrink the orbit indefinitely, ending with the two unsynchronised stars plunging into each other. This runaway process is called the \emph{Darwin instability}\footnote{The Darwin instability is named after astronomer and mathematician George H. Darwin, the fifth child of renowned English naturalist Charles Darwin.} \citep{darwin1879a}.
\\ \\
\noindent\textbf{Secular interactions in stellar multiples}
We have so far discussed binary stars in isolation. The influence of a distant third body in a so-called \emph{hierarchical triple system} can bring the inner binary into contact. This is particularly important for massive stars because stellar multiplicity is observed to increase with mass, with the majority of O-type \ac{MS} stars being in triple and quadruple systems \citep{moe2017a,offner2023a}. When the relative inclination between the outer orbit and inner binary orbit exceeds ${\approx}\,39^\circ$, periodic oscillations between the inner binary's eccentricity and inclination take place, called \emph{von Zeipel-Lidov-Kozai oscillations} \citep{vonzeipel1910a,lidov1962a,kozai1962a}. If large enough eccentricities are excited in the inner binary, the stars could come into contact at periapsis.
\\ \\
\noindent\textbf{Dynamical interactions in dense stellar environments}
In dense stellar environments like globular clusters and nuclear star clusters, mergers can also occur from direct collisions between stars. In addition, distant dynamical encounters could shrink the binary separation and lead to contact. Whether such encounters tend to widen or tighten the binary orbit depends on how tight the binary is and the velocity dispersion of stars in the cluster. For a cluster with mass $M_\textrm{cluster}$ and radius $R_\textrm{cluster}$, the velocity dispersion is given by $\sigma = (GM_\textrm{cluster}/R_\textrm{cluster})^{1/2}$. So-called \emph{hard binaries} have binding energies exceeding the typical kinetic energy of stellar perturbers, i.e., $GM_1M_2/a > m\sigma^2$, where $M_1$ and $M_2$ are the masses of the binary components, $a$ is the semi-major axis, and $m$ is the typical star mass in the cluster. On average, hard binaries further harden (decreasing their semi-major axes) while \emph{soft binaries} become softer, a result known as \emph{Heggie's law} \citep{heggie1975a}. For distant (secular) encounters, the orbital angular momentum changes much more efficiently than the energy. In this case, excitation of binary eccentricity can directly lead to contact at periapsis or activate other dissipative processes that shrink the orbit, such as tides.

\section{Stellar mergers}\label{chap1:stellar-mergers}

In this section, we consider stellar mergers as dynamical encounters where the structures of both stars are highly relevant and their respective material is mixed in the final merger product. Historically, stellar mergers were first considered in very dense stellar systems such as galactic nuclei and globular clusters where the repeated head-on collisions and supernova explosions from the collision products were suggested to power the then newly discovered but yet ill-understood quasars \citep[\eg][]{spitzer1966a, colgate1967a, mathis1967a, deyoung1968a}. Today, quasars are known to be supermassive black holes that shine bright in the radio regime thanks to ongoing accretion of matter, and stellar mergers are responsible for a plethora of other interesting stellar objects and transients. For example, some blue stragglers are merger products and the most massive stars in any stellar population are likely such objects \citep[\eg][]{hills1976a, schneider2014a}. Luminous red novae (sometimes also called gap transients; see also Sect.~\ref{subsec:LRNe}) are also connected to the process of merging \citep[][]{munari2002a, tylenda2011a} and maybe the same holds for the great eruption of $\eta$~Carinae \citep{smith2018a, hirai2021a}. Some magnetic massive stars and their highly magnetic remnants, the most magnetic white dwarfs (polars) and neutron stars (magnetars), may have obtained their strong magnetic fields in the merger process \citep{ferrario2009a, wickramasinghe2014a, schneider2019a}. Moreover, some remarkable supernovae such as SN~1987A \citep[\chapterreft{SN~1987A}; \eg][]{podsiadlowski1989a, podsiadlowski1990a} and interacting (superluminous) supernovae \citep[\eg][]{justham2014a} are likely from merged stars. Some merged stars will collapse into black holes and could explain very massive stellar-mass black holes observed in gravitational-wave merger events \citep[\eg][]{dicarlo2019a, renzo2020c, costa2022a, schneider2024a}. In the following, we will describe the basics of the stellar merger process, explain the structure of merged stars, and discuss their subsequent evolution and final fates.

\subsection{Simplified picture of merging stars: ``entropy sorting''}\label{chap1:mergers:entropy-sorting}

The merger process is highly complex, and we first develop a simplified picture of it. To this end, we apply the concept of buoyancy, \ie Archimedes' principle, to stars. Consider a fluid element inside a star that has a certain mass, temperature and pressure, and that is displaced upwards. The fluid element is assumed to achieve pressure equilibrium with the surroundings while conserving its entropy, \ie there is no heat exchange. Archimedes' principle then states that a stellar structure is stable against buoyancy if a displaced fluid element has a higher density than its surroundings, which implies it will sink back to its original position. Mathematically, one can show that a star is stable against buoyancy if \citep[see \eg][]{landaulifshitz6eng}
\begin{equation}
    \frac{\mathrm{d} s}{\mathrm{d} r} > 0,
    \label{eq:stability-convection-entropy}
\end{equation}
where $s$ is the specific entropy of a mass shell inside the star and $r$ is the radius. That is, the entropy inside a star increases outwards for a star which is stable against such fluid displacements. In stellar evolution theory, Eq.~(\ref{eq:stability-convection-entropy}) is typically written in terms of logarithmic temperature gradients, $\nabla \equiv (\mathrm{d} \log T/\mathrm{d} \log P)$, and is known as the Schwarzschild criterion for stability against convection \citep[see \eg][]{kippenhahn2012a}.

We now apply this concept to the structure of a stellar merger product, which has an outwardly increasing entropy profile once it has reached dynamic equilibrium. If the merger process was adiabatic, the entropy of fluid elements would be conserved, and we could then immediately predict the merged star's structure. The fluid elements with the lowest entropy sink to the centre of the merged star, while those with the highest entropy settle at the surface, and so on\footnote{Within the picture of buoyancy, low-entropy material behaves like a stone in water while high entropy material is more like a cork floating on water.}. However, entropy is not conserved during the merger process because of shocks and chemical mixing, and entropy sorting alone can thus not predict the structure of the merged star entirely correctly. In practice, methods have been developed that first modify the original entropy of the pre-merger stars according to knowledge gained from three-dimensional (3D) collision simulations\footnote{Often, one does not use entropy directly but an entropic variable that is mathematically very similar to entropy.} and only then compute the structure of the merged star following the idea of entropy sorting \citep[\eg][]{lombardi1996a, lombardi2002a, gaburov2008a}. Indeed, also some of the latest detailed 3D magnetohydrodynamic simulations of a stellar merger \citep{schneider2019a} show an outwardly increasing entropy profile in the merger product.

\subsubsection{Mergers of two MS stars versus mergers of a post-MS star and a MS star}\label{chap1:mergers:ms+ms-vs-postMS+MS-mergers}

\begin{figure}[tb]
    \centering
    \includegraphics[width=\textwidth]{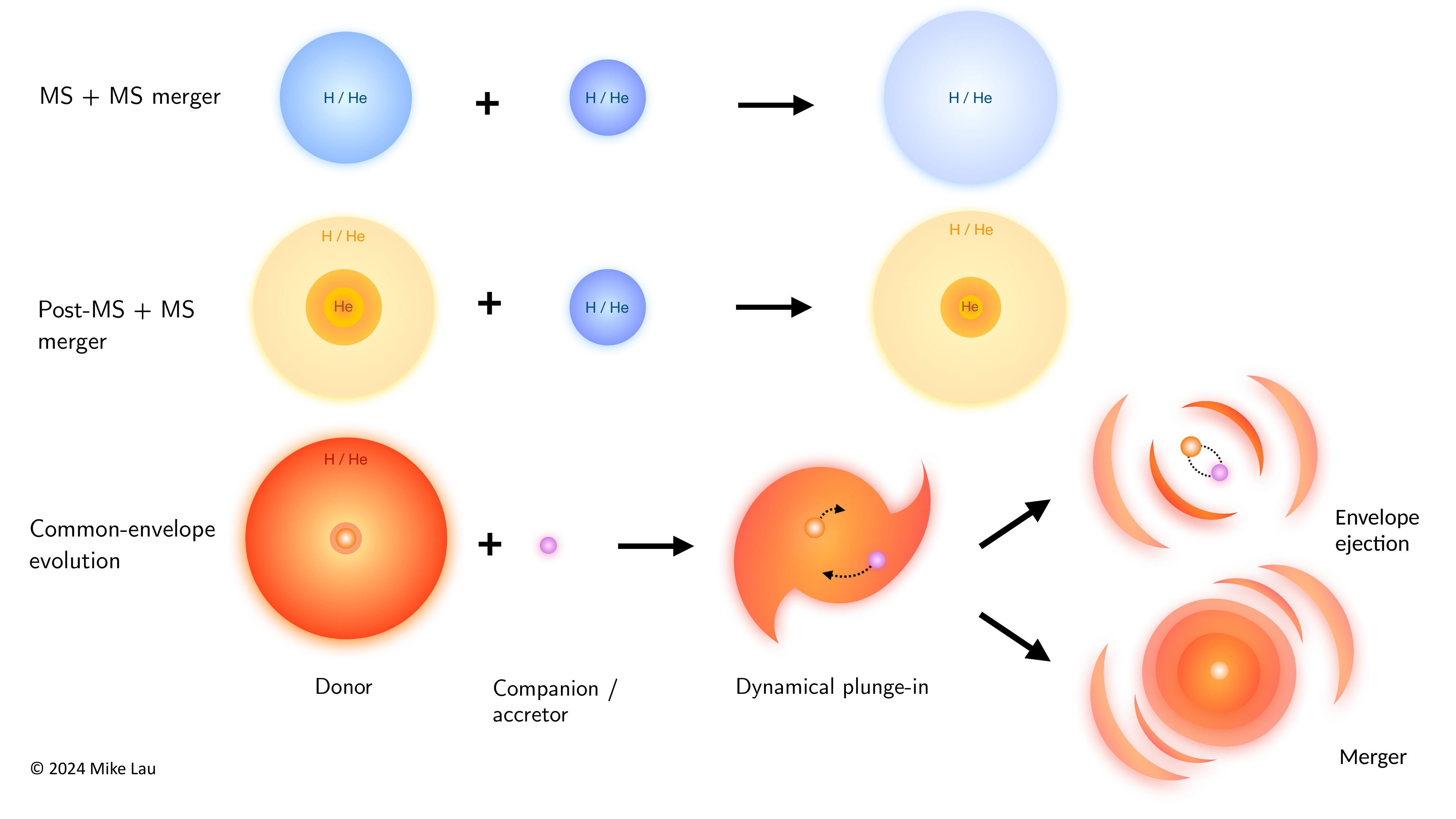}
    \caption{Illustration of stellar mergers and \ac{CE} evolution as processes initiated at different evolutionary phases of the primary star. In the main-sequence (MS) phase, hydrogen (H) is fused into helium (He) in the core. In post-MS evolution, the primary may have an inert helium core, a helium-burning core producing carbon and oxygen, or an inert carbon-oxygen core. Only the first case is illustrated here. Evolved primary stars can also act as donors in a \ac{CE} interaction, typically in later evolutionary stages where a clear separation between the core and the envelope exists. The outcome of the dynamical plunge-in can either be an ejection of the \ac{CE} or a merger of the stellar cores. Note that the objects are not drawn to scale.}
    \label{fig:merger-cartoon}
\end{figure}

To demonstrate the concept of entropy sorting and to gain further insight into the interior structure of merged stars, we now consider mergers of two MS stars and of a post-MS and a MS star; Fig.~\ref{fig:merger-cartoon} illustrates such mergers. MS stars are characterised by the nuclear fusion of hydrogen into helium in their cores and a hydrogen-rich envelope, where no nuclear burning takes place. In an MS+MS star merger, the hydrogen-burning cores have the lowest entropies and thus form the new core of the merged star. Some of the helium and other hydrogen-burning products are typically mixed out of the two cores and spread over the new envelope. In many cases, the newly formed star can fully adjust to its new structure such that it resembles a genuine single star of the same mass as the merged star (except for the additional chemical mixing). Whether a merger product can indeed adjust its structure to that of an equivalent single star depends crucially on a process called semi-convection and we refer the reader to \cite{braun1995a} for more details. If the adjustment is not possible, the merged star will have a smaller convective core and a larger hydrogen-rich envelope than a genuine single star of the same mass and same central hydrogen abundance.

In post-MS stars, there is no hydrogen in the core and fusion has ceased. Here, we consider the specific evolutionary phase where the core consists of helium, and hydrogen fusion takes place in a layer on top of it (see Fig.~\ref{fig:merger-cartoon})\footnote{There are also post-MS stars that have finished core helium burning and now have a carbon-oxygen core, surrounded by a helium-burning layer, a hydrogen-burning layer and a non-burning envelope. This is the typical structure of a red (super-)giant after core-helium burning and such structures are discussed in this chapter in the context of common-envelope evolution.}. The helium core has a much higher temperature and density than that of an MS star and usually a significantly lower entropy. Hence, the helium core will sink to the centre of the merged star. As in the case of the MS+MS star merger, a fraction of the helium core will be mixed with outer layers and thereby enrich them in helium and other hydrogen-burning products (\eg nitrogen). If the primary star is sufficiently evolved that its core already fuses helium into carbon and oxygen, such elements would also be mixed into the outer layers. In a post-MS+MS star merger, the remnant cannot adjust its structure to that of a genuine single star with the same mass. Hence, the mixing of parts of the helium core implies that the helium core of the merged star is less massive than that of the progenitor post-MS star. The hydrogen-burning shell of the post-MS star and the core of the MS star will surround the helium core and both are surrounded by the former hydrogen-rich envelopes of the two progenitor stars. Hence, in comparison to equivalent single stars, a post-MS+MS merger remnant has a smaller helium core mass and a larger hydrogen-rich envelope mass. Depending on the core-to-envelope mass ratio and the exact chemical abundances in these layers, such stars can burn hydrogen in a thick convective shell. These structures resemble the situation in which a massive MS star resides on top of a helium core. Massive MS stars have convective cores (the now thick convective hydrogen-burning shell) and radiative envelopes. Hence, such merger remnants look like blue supergiants and may even remain so for the rest of their evolution \citep[see \eg][]{hellings1983a, podsiadlowski1989a, braun1995a, claeys2011a, vanbeveren2013a, justham2014a, schneider2024a, menon2024a}.

\subsection{More complete picture of merging stars: exemplary magnetohydrodynamic simulation}\label{chap1:mergers:3d-mhd-example}

\begin{figure}[tb]
    \centering
    \includegraphics[width=1.0\textwidth]{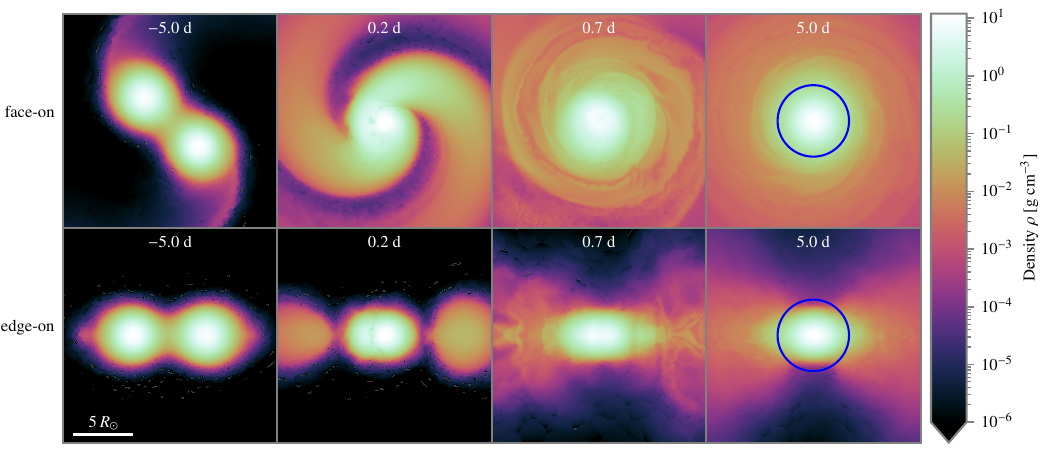}
    \includegraphics[width=1.0\textwidth]{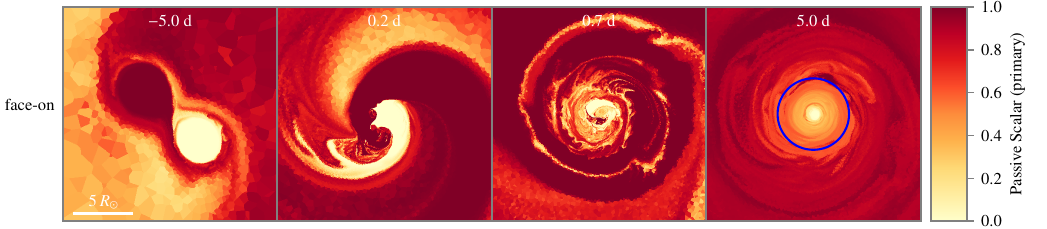}
    \caption{Evolution of mass density in the $x$--$y$ (face-on) and $x$--$z$ (edge-on) planes (top two rows), and the fraction of material originating from the former $9\,\msun$ primary star in the $x$--$y$ plane (passive scalar; bottom row). The assumed main merger is at a time $t=0\,\mathrm{d}$. The blue circle at $t=5\,\mathrm{d}$ with a radius of $3\,\rsun$ indicates the spherically symmetric central merger remnant which is surrounded by a massive torus/disc-like structure. Image courtesy of Sebastian Ohlmann.}
    \label{fig:mhd-merger-density-pass-scalar}
\end{figure}

For a more complete understanding of stellar mergers, we need to consider 3D (magneto-)hydrodynamic simulations, and we will consider the merger of initially $9$ and $8\msun$ MS stars from \citet{schneider2019a}. The stars are relatively unevolved and only finished about 30\% of their MS evolution. We show the density evolution of this merger in Fig.~\ref{fig:mhd-merger-density-pass-scalar} and also the fraction of material originating in the initially $9\msun$ primary star (labelled as ``passive scalar''). 

The two stars orbit each other and get closer because of mass being lost through the outer Lagrange points, removing angular momentum from the binary star. At $t=0.2\,\mathrm{d}$\footnote{The time $t=0\,\mathrm{d}$ is an ad-hoc definition and marks the moment when both stars truly merge into one object.}, both stars are tidally disrupted and coalesce into the merger remnant. Because of the angular momentum that is mostly from the initial binary orbit, the remnant consists of two main components: a central spherically symmetric merger remnant of about $14\,\msun$ and a disc-like, rotationally and pressure-supported torus of about $3\,\msun$. The torus holds the majority of the initial angular momentum (about 60\%) and is expected to be accreted within a few years onto the central remnant. As radiative cooling in this case is likely inefficient on such short timescales, the expectation is that most of the torus is transformed into an extended and heated envelope of the merger product \citep[][]{schneider2019a, schneider2020a}. Because of obstruction by the massive torus, material lost from the central merger remnant via, \eg, enhanced winds from the energy deposition of the merger process inside the central merger remnant or other processes will be along the $z$ direction, \ie perpendicular to the orbital plane. Hence, a bipolar nebula structure is expected to be present for some time after the merger until it has expanded so much that it cannot be observed anymore.

The ``passive scalar'' in Fig.~\ref{fig:mhd-merger-density-pass-scalar}, which is the fraction of material originating from the primary star, demonstrates the mixing taking place in the merger and further showcases the highly turbulent merger process. In this particular case, the core of the initially $8\,\msun$ secondary star sinks into the centre of the merger remnant while some core material of the initially $9\,\msun$ primary is deposited in the torus, which forms part of the future envelope of the merger remnant. This outcome is not expected from the idea of entropy sorting detailed in Sect.~\ref{chap1:mergers:entropy-sorting} that would predict that the core of the initially more massive $9\,\msun$ star forms the core of the merged star. This outcome is typical of almost equal-mass stellar mergers, while mergers with more discrepant masses do follow the general expectations from entropy sorting \citep[see \eg][]{glebbeek2013a}. This example is specifically chosen here to highlight a case where entropy sorting is expected to give a qualitatively wrong result.

In summary, the merger of two stars leads to their tidal disruption and mixing into a new structure. When there is enough angular momentum present as in the highlighted example, the immediate merger product consists of a centrally, spherically-symmetric remnant surrounded by a massive disc-like structure. In head-on collisions, there may not be any angular momentum such that there is no massive torus. This is a qualitatively very different outcome.

\subsection{Stellar merger debris}\label{chap1:mergers:ejecta-masses}

\begin{figure}[tb]
    \centering
    \includegraphics[width=0.6\textwidth]{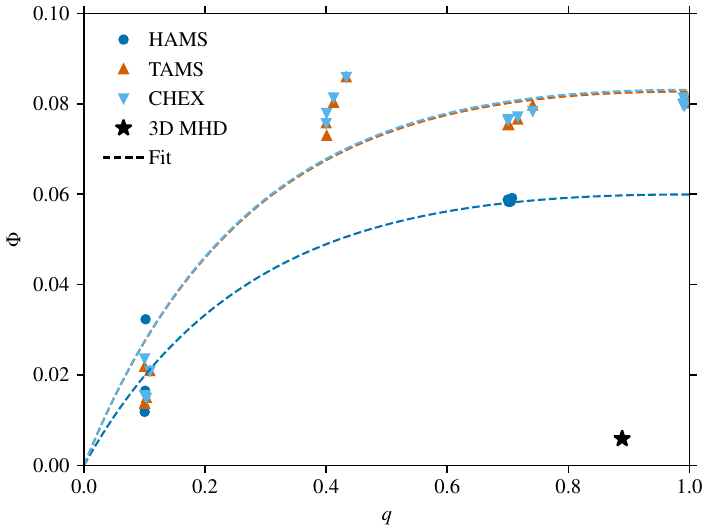}
    \caption{Fraction $\Phi=\Delta M/(M_1 + M_2)$ of dynamic mass loss as a function of mass ratio $q=M_2/M_1$ in the head-on collisions simulated by \citet{glebbeek2013a} and the circular binary merger simulated by \citet{schneider2019a}, denoted ``3D MHD''. Here, $\Delta M$ is the ejected mass and $M_1+M_2$ is the total mass of the binary systems before the merger. The collisions of \citet{glebbeek2013a} involve half-age MS (HAMS), terminal-age MS (TAMS) and core-hydrogen exhausted (CHEX) stars. The dashed lines fitted to the simulation data following Eq.~(\ref{eq:merger-dynamic-mass-loss}) have $C=0.240\pm0.012$, $C=0.331\pm0.011$ and $C=0.332\pm0.010$ for the HAMS, TAMS and CHEX mergers, respectively. Figure adapted from the Master's thesis of Max Heller, Heidelberg University, 2024.}
    \label{fig:merger-dynamic-mass-loss}
\end{figure}

In a merger, some material is ejected dynamically and leaves the system. Such material can form a nebula surrounding the merged star and, as described in Sect.~\ref{chap1:mergers:3d-mhd-example}, is likely bipolar because of preferential ejection along the axis perpendicular to the former orbital plane. The ejecta mass $\Delta M$ depends on the details of the mergers. In head-on collisions, a large range of ejecta fractions, $\Phi = \Delta M / (M_1 + M_2)$ (where $M_1$ and $M_2$ are the masses of the two stars before the merger and $M_1>M_2$ in our definition), is possible depending on the collision velocity and impact parameter. \citet{freitag2005a} find values of $10^{-5} \leq \Phi \leq 0.7$, where the smallest ejecta masses are obtained for the slowest collisions with the largest impact parameters such that the surfaces of both stars just touch each other. The largest values are found in direct head-on collisions with the fastest velocities. In the head-on collisions of two MS stars of \citet{glebbeek2013a}, the stars are initially on a parabolic orbit, \ie there is no orbital energy and angular momentum (the merged remnant will not rotate). These authors find that the ejecta fraction is well-described by
\begin{equation}
    \Phi = \frac{\Delta M}{M_1 + M_2} = C \frac{q}{(1+q)^2}
    \label{eq:merger-dynamic-mass-loss}
\end{equation}
with $C\,{\approx}\,0.3$ (see Fig.~\ref{fig:merger-dynamic-mass-loss}), \ie they range from a few to ten per cent. In these collisions, one can even see a dependence of the ejecta fraction on the evolutionary stage of the colliding stars. More evolved stars generally have larger ejecta fractions than less evolved stars because the binding energies of the stellar envelopes become smaller with later evolutionary stages. In the most extreme cases of giants and supergiants, the entire envelopes might be ejected (\ie a \ac{CE} interaction).

Contrarily, in the generally slower and smoother mergers of binaries with almost circular orbits as described in Sect.~\ref{chap1:mergers:3d-mhd-example}, the ejecta masses are smaller by about a factor of 10 (data point ``3D MHD'' in Fig.~\ref{fig:merger-dynamic-mass-loss}) and only $\approx 1\%$ of the total binary mass is lost in a $q\approx0.9$ merger of relatively unevolved MS stars. If one were to scale down the fit curve to the half-age MS data in Fig.~\ref{fig:merger-dynamic-mass-loss} to this one data point, the equivalent value of $C$ in Eq.~(\ref{eq:merger-dynamic-mass-loss}) would only be ${\approx}\,0.024$.

\subsection{Rejuvenation}\label{chap1:mergers:rejuvenation}

Stellar mergers of MS stars are known to produce so-called \textit{blue straggler stars} \chapterref{Blue stragglers}. Blue stragglers are stars that appear younger to an observer than they are---they are thus often said to have \textit{rejuvenated} \citep[see \eg][]{hellings1983a, podsiadlowski1992a, braun1995a, dray2007a, schneider2016a}. Rejuvenation has essentially two contributions: \textit{apparent} and \textit{intrinsic}/\textit{true rejuvenation}. The former refers to the fact that a merger of two stars naturally leaves behind a star with a higher mass and thus larger luminosity. Such stars can be situated beyond the MS turn-off in a star cluster\footnote{During the evolution of a population of coeval stars, \eg in a star cluster, the initially most massive stars end their MS evolution first and thereby disappear from the MS in the Hertzsprung--Russell diagram of all stars. This leads to a point on the MS, the MS turn-off, above which no stars are found. This turn-off can be used to determine the age of a star cluster as the MS lifetime of the presently most massive star in the cluster.}. Because a star's lifetime scales inversely with its mass, one would be led to the conclusion that a star beyond the MS turnoff is intrinsically younger than other stars in the cluster. Hence, one would attribute a too-young age to the star. The intrinsic or true rejuvenation refers to the fact that the central hydrogen fuel in a merged star can be lower than that of both parent stars because of the mixing of fresh hydrogen into the core of the merged star. This mixing partly happens during the dynamic merger process (see Fig.~\ref{fig:mhd-merger-density-pass-scalar}) and it can additionally occur in the evolution of the merged star right after the coalescence when it re-establishes thermal equilibrium and transient convective cores have been found \citep[see \eg][]{schneider2020a}. In a stellar evolution context, one often speaks of merged stars having fully rejuvenated if their interior structures were able to adjust to that of a genuine single star of the same mass. Whether this is possible at all, is linked to the so-called semi-convection and its efficiency of mixing through layers with a strong chemical gradient \citep[see][for more details]{braun1995a}. 

\citet{glebbeek2008a} developed a formalism that predicts the amount of rejuvenation and the apparent age of merged MS stars. They calibrated the model against the outcome of head-on collision simulations of low-mass stars. For a calibration of the same model to high-mass head-on collision products, see \citet{glebbeek2013a} and, for a version of the same model extended to also account for stellar wind mass loss of massive stars, we refer the reader to \citet{schneider2016a}.

In all cases, the difference between true and apparent age, \ie the rejuvenation, can be significant. It ranges from at least a few per cent up to 80\% and more of the MS lifetime of stars, and depends on the masses of the merged stars and their evolutionary stage \citep{schneider2016a}. This implies that blue stragglers from merged stars could appear up to a factor of about 10 younger than other stars formed at the same time \citep[see also Fig.~16 in][]{schneider2015a}.

\subsection{Magnetic field amplification in stellar mergers}\label{chap1:mergers:b-fields}

\begin{figure}[tb]
    \centering
    \includegraphics[width=1.0\textwidth]{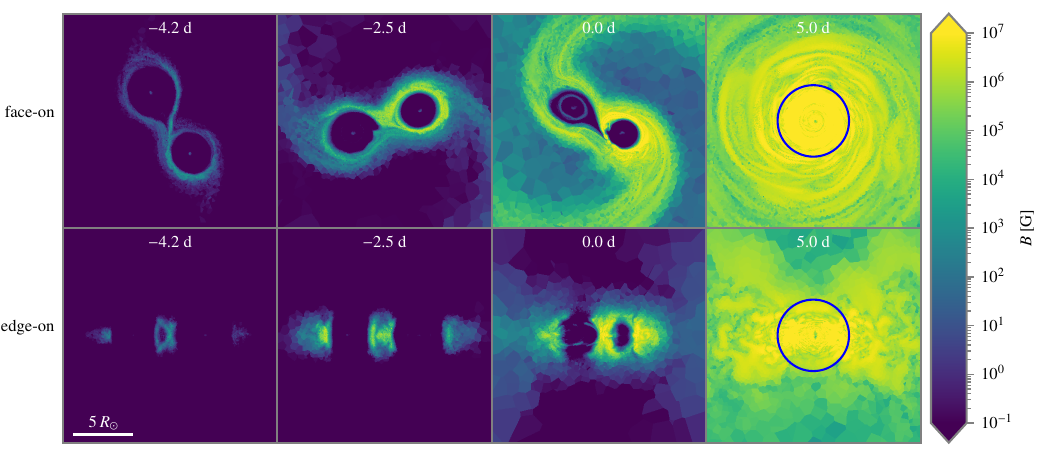}
    \caption{Evolution of the absolute magnetic field strength of the MS+MS $9$ + $8\,\msun$ star merger from \citet{schneider2019a} in the face-on and edge-on view (see Fig.~\ref{fig:mhd-merger-density-pass-scalar} for the evolution of the mass density and the mixing of the two stars during the merger process). As before, time $t=0\,\mathrm{d}$ corresponds to the actual merger of the two stars and the circle indicates the approximate size of the central spherically-symmetric merger remnant with a radius of about $3\,\rsun$. Image courtesy Sebastian Ohlmann.}
    \label{fig:mhd-merger-abs-bfield}
\end{figure}

The merger of two stars is a highly turbulent process in which efficient magnetic field amplification is expected. The 3D magnetohydrodynamic simulations by \cite{schneider2019a} show that a negligible seed magnetic field is rapidly amplified in the $9$ + $8\,\msun$ MS star merger described in Sect.~\ref{chap1:mergers:3d-mhd-example} (Fig.~\ref{fig:mhd-merger-abs-bfield}). Various mechanisms and instabilities are acting together to produce the highly magnetised merger remnant. In the beginning, \eg at $t=-4.2\,\mathrm{d}$ in Fig.~\ref{fig:mhd-merger-abs-bfield}, shear instabilities and a continuous winding-up of the seed magnetic field amplify the field until the magneto-rotational instability \citep{balbus1991a, balbus1995a} is resolved in the simulation, which greatly and further helps the amplification process. The first phase of amplification mainly takes place in the accretion stream from the primary to the secondary star, particularly in the shear layer that it creates when the flow surrounds the secondary and later also the primary star. In the actual merger, Kelvin-Helmholtz instabilities are seen prominently at the interface of the two merging stars, which further help to amplify the magnetic field. After the actual merger of the two stars at $t=0\,\mathrm{d}$, a large-scale dynamo sets in that is responsible for generating a more ordered, larger-scale magnetic field from the small-scale magnetic fields produced by the processes mentioned above. The magnetic field amplification processes in a large variety of binary star mergers are very similar and we refer to \citet{schneider2019a} for mergers of MS stars, to \citet{pakmor2024a} for white dwarf mergers, to \citet{kiuchi2024a} for neutron star mergers, and to Sect.~\ref{chap1:ce-bfields} for \ac{CE} events.

The amplification processes stop and an asymptotic magnetic energy is reached regardless of the exact initial conditions of the seed magnetic field and orbital configuration of the simulations \citep[\eg][]{schneider2019a, ondratschek2022a, pakmor2024a, kiuchi2024a}. From a theoretical point of view, this is the expected behaviour when energy equipartition is reached between the magnetic field and those energy reservoirs from which it is fed. In a simplified phenomenological model, the source of magnetic energy is turbulent energy described by a characteristic turbulent velocity $v_\mathrm{turb}$ such that, upon energy equipartition of the turbulent $e_\mathrm{turb}$ and magnetic $e_\mathrm{B}$ energy densities (\ie $e_\mathrm{turb}\,{\simeq}\, e_\mathrm{B}$), we have
\begin{equation}
    \frac{1}{2}\rho v_\mathrm{turb}^2 \simeq \frac{B^2}{8\pi},
    \label{eq:energy-equipartition}
\end{equation}
where $B$ is the absolute magnetic field strength. Assuming that $v_\mathrm{turb}$ is related to differential rotation and hence the Keplerian (orbital) velocity, $v_\mathrm{Kep}^2\,{=}\,GM/R$, of a merger remnant with mass $M$ and radius $R$, we find (assuming $v_\mathrm{turb} \,{\sim}\, v_\mathrm{Kep}$ and setting $\rho \,{\sim}\, M/R^3$)
\begin{equation}
    B \simeq 10^3\,\mathrm{G} \: \left( \frac{M}{5\,\msun} \right) \left( \frac{3\,\rsun}{R} \right)^{2}.
    \label{eq:merger-b-field}
\end{equation}
Such a simple model correctly predicts a constant magnetic flux per unit mass ($B R^2/M = \mathrm{const.}$) for merger-induced magnetic fields as is observationally inferred by \citet{wickramasinghe2014a} for magnetic massive MS stars and highly magnetic white dwarfs. Eq.~(\ref{eq:merger-b-field}) is in rough agreement with the known magnetic fields of magnetic massive stars (${\sim}\,10^3\,\mathrm{G}$), highly magnetic white dwarfs (polars, ${\sim}\,10^{8}\,\mathrm{G}$), and highly magnetic neutron stars (magnetars, ${\gtrsim}\,10^{12}\,\mathrm{G}$). Hence, these objects could have obtained their magnetic fields in a similar process.

\subsection{Rotation and surface chemical enrichment of merger products}\label{chap1:mergers:rotation+surface-enrichment}

In binary star systems, the orbit usually contains more angular momentum than is possible to have in rotating stars. Hence, the naive expectation is that merged stars must be rotating very rapidly. However, this picture is challenged by the works of \citet{schneider2019a, schneider2020a} and earlier ideas of \citet{leonard1995a}. Solid-body rotation is the minimum energy state of a (radially) differentially rotating star. In the simulations shown in Fig.~\ref{fig:mhd-merger-density-pass-scalar} and~\ref{fig:mhd-merger-abs-bfield}, the merger product approaches this state within $5\text{--}10\,\mathrm{d}$ after the coalescence. Assuming solid-body rotation and a surface spinning at the critical Keplerian velocity for the merger product after accreting the torus material, the puffed-up star only holds a few per cent of the initial orbital angular momentum of the binary progenitor system. Hence, when the star is back on the main sequence, it is a slow rotator. This implies that during the accretion of the massive torus, most of the angular momentum must be transported away. Hence, \citet{schneider2019a, schneider2020a} suggest that the co-evolution of the central merger remnant and the massive torus sets the initial angular momentum of a merged star. A simulation by \citet{schwab2012a} of a central white-dwarf merger remnant accreting from a massive torus supports this idea.

While the co-evolution of the central merger remnant and its massive torus may be able to transport away a significant fraction of the angular momentum, there are further processes that can quickly spin down the merged star. First of all, the merged star is highly magnetised such that magnetic braking may be efficient \citep[see also][]{leonard1995a}. Second, the merged star may undergo a thermal relaxation process in which it tries to radiate away excess heat from the merger process. This may lead to large expansion, evolution close to the Eddington luminosity and possibly strongly enhanced stellar winds---a combination that can remove a significant amount of angular momentum. Thirdly, magnetic disc-locking, outbursts and bipolar, jet-like outflows may be expected when accreting the torus as is observed in star formation and is known to regulate the angular momentum budget of newly formed stars \citep{hartmann2016a}. In conclusion, it is still unclear how fast merged stars rotate from a theoretical standpoint but several lines of thought suggest that such objects can be slow rotators. Observational clues to the rotation of merger products will be discussed in Sect.~\ref{chap1:mergers:examples-properties}.

A similarly uncertain question is whether the surface chemical abundances of merged stars show any anomalies. As detailed in Sect.~\ref{chap1:mergers:entropy-sorting}, there is the mixing of former core material into the envelope of the merged stars. This mixing will dredge up chemical elements that were processed by nuclear burning in the cores of stars. At the moment, it is unclear whether this mixing extends up to the surface of the merged star or whether it remains hidden in deeper layers of the envelope of the merged star. If it becomes visible, one expects to find enhanced helium and nitrogen surface abundances as well as a depletion of carbon and oxygen as a result of hydrogen burning via the CNO cycle. For hydrogen burning via the pp-chains, one may expect enhanced $^3\mathrm{He}$ abundance besides the obvious enrichment of $^4\mathrm{He}$ (plus some nitrogen enhancement because of slow CN(O) cycling) and lithium depletion. Generally speaking, mixing is found to be more efficient for more massive merged stars and more evolved progenitors \citep[\eg][]{lombardi1996a, sills1997a, glebbeek2013a}.

\subsection{Post-merger evolution and final fates}\label{chap1:mergers:post-merger-evolution}

The immediate post-merger evolution is characterised by thermal relaxation of the merged star, during which the star radiates away excess energy that was deposited in the merger process. During this relaxation, the star may expand greatly, possibly to the extent that it becomes a fully convective (super)giant for a short amount of time. The excess energy may drive enhanced stellar winds that can potentially become super-Eddington in very massive merger products. After a global thermal timescale of the merger product, the star continues its further evolution. In MS+MS star mergers, the star would continue as a MS star and would likely be visible as a blue straggler in a star cluster \chapterref{Blue stragglers}. Post-MS+MS star mergers either continue their evolution as long-lived blue (super)giants or quickly become cool (super)giants depending on the exact post-merger interior structure (see Sect.~\ref{chap1:mergers:entropy-sorting}). In all cases, merged stars are likely overluminous compared to single stars because of the mixing of helium from the cores into the envelopes. This increases their overall mean molecular weight $\mu$ and thereby the luminosity because of the approximate stellar mass-luminosity relation $L\propto M^3 \mu^4$ \citep{kippenhahn2012a}.

\begin{figure}[tb]
    \centering
    \includegraphics[width=0.8\textwidth]{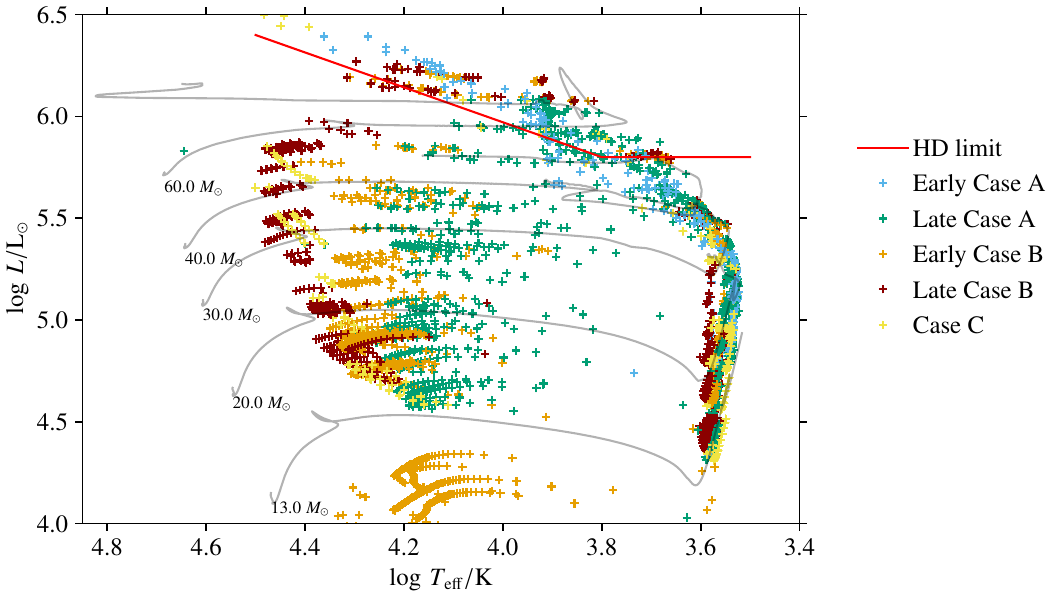}
    \caption{Hertzsprung--Russell diagram of the post-MS evolution of massive stellar merger products. Cross symbols are shown every $10^5\,\mathrm{yr}$ in the post-MS evolution of MS+MS mergers and post-MS+MS mergers. The grey thin lines are single-star evolution models for the indicated initial masses. Cases~A, B and C refer to mergers with ever more evolved primary stars: Case~A is for core-hydrogen burning, Case~B for post-core-hydrogen but pre-core-helium burning, and Case~C for post-core-helium burning primaries. The early Case~A merger products evolve much like single stars and quickly cross the Hertzsprung--Russell diagram without adding to the cool-supergiant population. The red solid line is the Humphreys--Davidson limit beyond which no stars are observed \citep{humphreys1979a}. The lack of models at the lower luminosity end in Fig.~\ref{fig:hrd-saling} is because of a lack of computed models and does not reflect a physical origin. Figure adapted from the Bachelor's thesis of Julian Saling, Heidelberg University, 2023.}
    \label{fig:hrd-saling}
\end{figure}

We illustrate the post-MS evolution of merger products in the Hertzsprung--Russell diagram in Fig.~\ref{fig:hrd-saling}. The merger models shown here are obtained by rapid mass accretion onto stars of various evolutionary stages and are as such quite simplified yet useful models. The post-MS evolution is separated into two broad classes: blue and cool (yellow and red) supergiants \citep{hellings1983a, podsiadlowski1989a, braun1995a, claeys2011a, vanbeveren2013a, justham2014a, schneider2024a, menon2024a}. Post-MS+MS star mergers and mergers with a very evolved MS star can give rise to a population of long-lived blue supergiants with effective temperatures $T_\mathrm{eff}\gtrsim 10,000\,\mathrm{K}$. Such long-lived blue supergiants are only found in a subset of mergers that result in stars with critically small core masses compared to their envelope masses (\cf Sect.~\ref{chap1:mergers:ms+ms-vs-postMS+MS-mergers}). Such stars may undergo supernova explosions as blue supergiants, which is not expected from classical single star evolution. The other merger products usually become cool supergiants that can have more massive envelopes compared to supergiants from genuine single stars. Blue supergiants from mergers offer a promising way to explain the large number of observed blue supergiants that are otherwise difficult to understand from single-star evolution \citep{bernini-peron2023a, menon2024a}. They may also give rise to unusual and extreme supernovae such as SN~1987A, interacting and superluminous supernovae \citep{podsiadlowski1989a, podsiadlowski1990a, justham2014a, menon2017a, schneider2024a}, and may form some of the most massive black holes \citep{dicarlo2019a, renzo2020c, costa2022a, schneider2024a}. Moreover, all merger products may leave behind strongly magnetised compact remnants such as highly magnetic white dwarfs (polars) and neutron stars (magnetars) if they can retain sufficient magnetisation from the merger process until the end of their lives.

\subsection{Examples and properties of merger products}\label{chap1:mergers:examples-properties}

Blue stragglers are the best-known and most-studied examples of merger products. However, merging stars is not the only way to produce blue stragglers, and binary mass transfer is an alternative. Apparently-single blue-straggler stars may thus be some of the best merger candidates. Interestingly, there is observational evidence that blue stragglers formed from merging may indeed be rather slow rotators as discussed in Sect.~\ref{chap1:mergers:rotation+surface-enrichment} \citep{wang2022a, ferraro2023a}. Further studying these stars will undoubtedly allow for new insights into the outcome of stellar mergers. When massive blue stragglers ($\gtrsim 2\,\msun$, \ie those that ignite helium not under degenerate conditions) age and evolve into red (super)giants, they will be visible as so-called ``red stragglers'' in coeval stellar populations \citep{britavskiy2019a}, \ie the most luminous red (super)giants in stellar clusters will be the descendants of former blue stragglers while the true single stars are among the least luminous red (super)giants. 

Other good merger candidates are blue (super)giants as explained in Sects.~\ref{chap1:mergers:ms+ms-vs-postMS+MS-mergers} and~\ref{chap1:mergers:post-merger-evolution}. But what are the best fingerprints of past stellar mergers and how can one identify such stars observationally? 
\newline
\begin{description}
    \item[Rejuvenation/age discrepancy] This is arguably the best technique to find stars that have accreted mass either by binary mass transfer or a stellar merger. To find such stars, a comparison clock is needed to which the apparent age of the merger candidate can be compared. Comparison clocks can be another gravitationally bound star in a wide orbit such that binary mass transfer can be excluded as the cause of the age discrepancy and/or a coeval host star cluster or stellar association.
    \item[Unusual rotation] While it remains uncertain how fast/slow merger products rotate, most of them may either rotate slowly or rapidly as discussed in Sect.~\ref{chap1:mergers:rotation+surface-enrichment}. There is evidence that at least a good fraction of merger candidates rotate slowly.
    \item[Chemical surface abundances] The mixing of chemical elements processed by nuclear burning from the core to the surface in a merger can enrich the surface in, \eg helium and nitrogen, and may deplete the surface in, \eg carbon and oxygen (see Sect.~\ref{chap1:mergers:rotation+surface-enrichment}). Similar signatures are expected from binary mass transfer and rotationally induced mixing.
    \item[Strong and large-scale surface magnetic fields] Stellar mergers provide the ideal environment for amplifying magnetic fields. If these magnetic fields are long-lived and organise themselves in large-scale structures, they may be observable on the surface of merger products for a good fraction of their lifetimes. 
    \item[Circumstellar nebula] For a short amount of time (some $10^4\,\mathrm{yr}$) a nebula may be visible surrounding a recently merged star until it has expanded so much that it cannot be observed anymore. Such nebulae are expected to be bipolar.  
    \item[Asteroseismology] Stellar mergers can have distinctly different interior structures (\eg small cores, thick hydrogen-burning shells, unusual chemical gradients inside stars, overall helium-enriched envelopes etc.) that lead to oscillation properties in p- and g-mode pulsators that cannot be explained by single-star models \citep{rui2021a, deheuvels2022a, li2022a, bellinger2024a, henneco2024b, wagg2024a}.
\end{description}

\hfill \newline
None of these properties is an exclusive sign of a past stellar merger but they may serve as evidence to support a merger hypothesis. For example, the stars $\tau$~Sco and HR~2949 have been identified as likely merger remnants because of an age discrepancy and their strong surface magnetic fields \citep{schneider2016a}. Another such example is the magnetic massive star HD~148937 that has a bipolar nebula and for which \citet{frost2024a} were able to find a binary companion in a wide orbit. The pair shows an age discrepancy that can be explained by a past merger. This star is a fast rotator, indicating not enough time has elapsed since the merger for it to be spun down. The latter star is arguably the most convincing merger candidate that links strong magnetism to stellar mergers. Other examples are the so-called B[e] stars where the forbidden emission lines indicate a nearby circumstellar nebula (\eg R4 in the Small Magellanic Cloud and the Galactic FS CMa star IRAS~17449+2320), and $\eta$~Carinae whose giant eruption may be linked to the merger process that then produced the spectacular Homunculus nebula \citep{smith2018a, hirai2021a, dvorakova2024a}.

\section{Common-envelope evolution}\label{chap1:cee}
\ac{CE} evolution was originally born out of theoretical necessity to explain the existence of evolved, short-period binaries. An example is V471 Tau, an eclipsing white-dwarf MS binary with a 12-hour period (separated by 3.2\rsun) that motivated the description of \ac{CE} evolution by \cite{paczynski1976a}, who is widely credited with its proposal\footnote{In his seminal paper, Bohdan Paczy\'{n}ski acknowledges a number of sources of inspiration, including discussions with Jeremiah P. Ostriker (1973) and the PhD thesis of Ronald Webbink \citep{webbink1975a}.}. Such binaries pose what is commonly called the \emph{separation problem}, which is the puzzling observation that their presently measured periods of the order of days could not possibly accommodate the sizes of their progenitors on the giant branch. For V471 Tau, the orbital separation must be a few hundred times larger to accommodate the asymptotic giant branch star that shed its envelope to leave behind the white dwarf.

\begin{figure}[tb]
    \centering
    \includegraphics[width=0.5\textwidth]{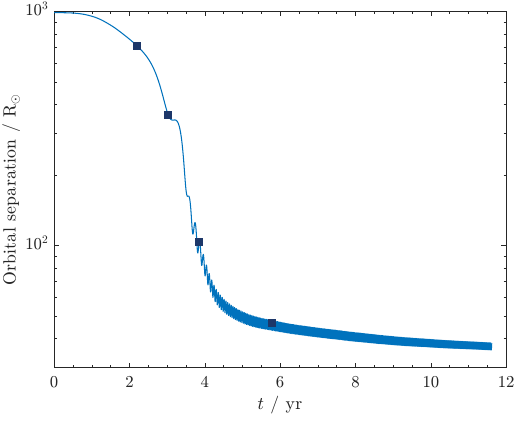}
    \caption{Separation between the donor core and the companion in a 3D hydrodynamical simulation of \ac{CE} evolution \citep{lau2022a}. The four square markers correspond to the snapshots presented in Fig. \ref{fig:3d_CE}.}
    \label{fig:sep}
\end{figure}

This problem is present in many other astronomical observations, including cataclysmic variables, X-ray binaries, hot subdwarf stars, Type Ia supernovae, and gravitational-wave driven mergers of binary black holes. All of them involve evolved stars or compact objects, expected to have experienced giant-branch evolution, in a tight orbit with another star. Explaining the origin of these objects requires understanding the \ac{CE} episode that might have drastically shrunk their orbital periods, among all binary interaction they might have experienced. The basic picture of \ac{CE} evolution is illustrated in the bottom row of Fig.~\ref{fig:merger-cartoon}. A compact companion star or accretor enters the extended envelope of an evolved giant star, the donor, following one of the pathways described in Section~\ref{chap1:contact-binaries}. A \emph{dynamical plunge-in} or \emph{spiral-in} ensues due to drag forces transferring energy and angular momentum from the orbit to the \ac{CE}. The result is that the envelope may become unbound, leaving behind a binary system formed from the core of the donor and the companion in a much tighter orbit. Fig.~\ref{fig:sep} shows the evolution in the orbital separation for this case, taken from a detailed \ac{CE} simulation. The dynamical plunge-in stops between the third and the fourth markers, after which the orbital separation changes relatively slowly. However, the envelope may remain bound if there is not enough energy, in which case the enduring drag causes the stellar cores to continue spiralling in until merging inside the \ac{CE}.

To this day, direct observational counterparts to the short-lived \ac{CE} phase remain scarce, and \ac{CE} evolution is still a largely theoretical concept. Combined with the difficulty in modelling this phase in detail, there remain many gaps in our understanding. Existing efforts have focused on performing one-dimensional (1D) and 3D hydrodynamical simulations of \ac{CE} and comparing observed populations of compact binaries to predictions based on simplistic prescriptions for the outcome of \ac{CE} evolution.

\subsection{Energy formalism and common-envelope efficiency, $\alpha_\mathrm{CE}$}

A rudimentary but instructive model of \ac{CE} evolution is given by the \emph{energy formalism} \citep{vandenheuvel1976a,webbink1984a}. It assumes that orbital energy does mechanical work on the \ac{CE} and causes it to become unbound, and so the binding energy of the envelope can be equated with the change in the orbital energy,
\begin{align}
    E_\mathrm{bind} = -\alpha_\mathrm{CE}\Delta E_\mathrm{orb},
    \label{eq:alpha}
\end{align}
where $\alpha_\mathrm{CE}$ is the \emph{common-envelope efficiency} that parametrises imperfect transferral of orbital energy into mechanical work. A minus sign is needed because we define $E_\mathrm{bind}$ as a positive quantity while the change in orbital energy is always negative. The expression for $\Delta E_\mathrm{orb}$ is  
\begin{align}
    \Delta E_\mathrm{orb} = \frac{GM_1 M_2}{2a_i} - \frac{GM_\mathrm{1,core}M_2}{2a_f} < 0,
    \label{eq:Eorb}
\end{align}
where $M_1$ is the donor mass, $M_2$ is the companion/accretor mass, $M_\mathrm{1,core}$ is the donor's core mass, $a_i$ is the orbital separation at the onset of the \ac{CE} event, and $a_f$ is the orbital separation after the \ac{CE} in-spiral. While $\alpha_\mathrm{CE}$ is usually an order-unity parameter, its exact range and its interpretation depend on how $E_\mathrm{bind}$ is defined. At its most basic form, $E_\mathrm{bind}$ is the gravitational binding energy of the \ac{CE}, that is, the work required to disperse the envelope to infinity. $E_\mathrm{bind}$ is therefore formally a positive quantity, although it is also frequently defined as the envelope's gravitational potential energy, in which case it is negative and no minus sign appears in Eq.~(\ref{eq:alpha}). It is also common to parameterise additional sources of energy that may aid envelope removal in $E_\mathrm{bind}$, which would decrease its value. For example, the binding energy of a 1D stellar structure can be calculated as
\begin{align}
    E_\mathrm{bind} = \int_M^{m_\mathrm{core}}\Bigg[-\frac{Gm}{r(m)} + \alpha_\mathrm{th}u(m) \Bigg]~dm.
    \label{eq:ebind}
\end{align}
The first term in the integrand is the gravitational potential at the mass coordinate $m$. The second term includes internal energy, $u$, as an additional energy source that is used with an uncertain efficiency, $\alpha_\mathrm{th}$. The idea is that gas thermal energy may be converted into mechanical work. Sometimes, \emph{ionisation potential} is also included in $u$. This is the potential energy per unit mass that is released if the stellar plasma cools enough that its constituent ions and electrons \emph{recombine} to form neutral atoms. The majority of stellar plasma is composed of ionised hydrogen, which releases \emph{recombination energy} of 13.6 eV per ion upon cooling to ${\sim}\num{6000}\,\mathrm{K}$.

If orbital energy were the only source of energy, $\alpha_\mathrm{CE}$ could not possibly exceed one. A value equal to unity is also unrealistic, because it implies perfectly efficient usage of orbital energy to unbind the envelope. That is, there are no radiative losses, and the ejected \ac{CE} envelope is exactly unbound, possessing zero kinetic energy at infinity. However, any energy source in addition to orbital energy that is not included in $E_\mathrm{bind}$ would effectively increase $\alpha_\mathrm{CE}$, even exceeding one. Examples include energy released from accretion by the stellar cores, which could be in the form of jets and kinetic outflows \citep{soker2015a,shiber2017a,shiber2019a}. If $E_\mathrm{bind}$ is defined without the internal energy contribution in Eq. (\ref{eq:ebind}), the conversion of recombination energy and thermal energy into mechanical work will also manifest as a larger $\alpha_\mathrm{CE}$. It is therefore important to understand the definition of $E_\mathrm{bind}$ when interpreting values of $\alpha_\mathrm{CE}$ presented in the literature.

\subsection{Gravitational drag}
\label{subsec:drag}

While we have mentioned orbital energy as a key energy source involved in \ac{CE} ejection, we have yet to elaborate on the nature of the drag force that transfers this energy into mechanical work that expands the \ac{CE}. For stars and compact objects, drag exerted from direct ram pressure is usually irrelevant. Most of the time, the dominant drag force is \emph{gravitational drag}, which is caused by gravitational attraction between the embedded object and the overdense wake and Mach cone trailing it \citep{chandrasekhar1943a,ostriker1999a}. To first order, neglecting dependences on the Mach number, the gravitational drag force can be written as the product of the interaction cross-section and the momentum flux,
\begin{align}
    F_\mathrm{drag} \approx \pi R_\mathrm{eff}^2 \rho \Delta v^2,
    \label{eq:Fdrag}
\end{align}
 where $R_\mathrm{eff}$ is the effective interaction radius, $\rho$ is the local ambient density, and $\Delta v$ is the relative speed between the object and its background. For gravitational drag, the interaction radius is the radius within which the gravitational potential energy due to the embedded object becomes significant compared to the incoming material's kinetic energy. This is approximately the \emph{Bondi-Hoyle-Lyttleton} radius, $R_\mathrm{BHL}\,{=}\, 2GM_2/(\Delta v^2 + c_s^2)$, where $M_2$ is the mass of the embedded object and $c_s$ is the local sound speed. Substituting $R_\mathrm{BHL}$ into Eq. (\ref{eq:Fdrag}), one finds that $F_\mathrm{drag}\,{\propto}\, \Delta v^{-2}$ in the strong-shock limit ($\Delta v/c_s \,{\gg}\, 1$). Despite the momentum flux increasing with $\Delta v$, the interaction radius decreases with $\Delta v$, because the motion of faster material would only be significantly altered deeper inside the embedded mass's potential. Eq. (\ref{eq:Fdrag}) also approximately gives the ram-pressure drag force when replacing $R_\mathrm{eff}$ with the physical radius, $R_2$, of the embedded object. Stars and compact objects are usually sufficiently compact that $R_2 \,{\ll}\, R_\mathrm{eff}$, indicating that ram-pressure drag is not important. For substellar objects like planets and brown dwarfs, it could be the opposite. Ram-pressure drag may also become more important for stars and compact objects moving deeper inside the \ac{CE}, where the orbital speed is much higher than at the surface.

With Eq. (\ref{eq:Fdrag}), one can estimate the duration of the \ac{CE} spiral-in as the ratio of the orbital energy to the rate of energy dissipation via drag. To obtain approximate scalings (ignoring constants of order unity), we write the orbital energy as $E_\mathrm{orb} \,{\sim}\, GmM_2/a$ where $M_2$ is the accretor/companion mass and $m$ is the mass interior to the companion's current orbital separation, $a$. The drag luminosity is $L_\mathrm{drag}\,{\sim}\, F_\mathrm{drag} \Delta v$ where the relative speed can be approximated as the Keplerian speed, $\Delta v \sim [G(m+M_2)/a]^{1/2}$. Combining this with Eq. (\ref{eq:Fdrag}) with the Bondi-Hoyle-Lyttleton radius $R_\mathrm{eff} = R_\mathrm{BHL}$, we obtain
\begin{align}
    t_\mathrm{insp} \sim \frac{E_\mathrm{orb}}{L_\mathrm{drag}}
    \sim \frac{m+M_2}{M_2} \frac{\langle\rho_d\rangle}{\rho} t_\mathrm{dyn}\hspace{1.5em}\textrm{(Gravitational drag),}
\end{align}
where $\langle\rho_d\rangle = m/a^3$ is the mean donor density interior to the companion's position, $\rho$ is the envelope density at the companion's current position, and $t_\mathrm{dyn} \, {\sim}\, \{a^3/[G(m+M_2)]\}^{1/2}$ is the free-fall timescale or orbital period at the separation $a$. This demonstrates that the drag-mediated inspiral indeed proceeds on the dynamical timescale, which is of the order of the orbital period. Similarly, for ram-pressure drag, using $R_\mathrm{eff} \,{=}\, R_2$ instead gives
\begin{align}
    t_\mathrm{insp} 
    \sim \frac{m}{m+M_2} \frac{R_2}{a} \frac{\langle\rho_2\rangle}{\rho}~t_\mathrm{dyn}\hspace{1.5em}\textrm{(Ram-pressure drag),}
\end{align}
where $\langle \rho_2 \rangle = M_2/R_2^3$ is the companion's mean density. Crucially, when the background medium is sufficiently denser than the companion such that $\langle\rho_2\rangle /\rho \lesssim R_2/a$ (assuming $M_2\lesssim m$), the companion's azimuthal velocity rapidly damps and the object plunges in quickly instead of spiralling in over many pseudo-Keplerian orbits. Such is the case for the engulfment of a Jupiter-like planet ($M_2\sim10^{-3}\msun$, $R_2\sim 0.1\rsun$, $\rho \sim \msun/\rsun^3$, $a\sim\rsun$) by a solar-like star, where $t_\mathrm{insp} \sim 0.1 t_\mathrm{dyn}$.

\subsection{Simulations of common-envelope evolution}
Modelling \ac{CE} evolution in detail is notoriously challenging. The first difficulty lies in the vast range of spatial and temporal scales spanned by the physical processes involved in \ac{CE} evolution. Taking a more extreme example, the \ac{CE} phase involved in the formation of a neutron-star X-ray binary or a double neutron star has an initial orbital separation of a few astronomical units between a red (super)giant donor and a $\approx 10\km$ neutron star. This gives rise to a dynamic range spanning 7 to 8 orders of magnitude. The range of temporal scales is also vast. At the onset of dynamically-unstable Roche-lobe overflow, the plunge-in may only commence after hundreds of orbits. Yet, to maintain numerical stability, a typical hydrodynamical simulation cannot take timesteps larger than the sound-crossing time of the smallest resolution length scale. A second difficulty is the inherent 3D nature of the \ac{CE} inspiral, meaning simplifications to 1D and 2D are accompanied by many unrealistic assumptions. Finally, while gravity and hydrodynamics are the basic drivers of \ac{CE} evolution, many additional processes are potentially relevant, like magnetohydrodynamics, radiation transport, and dust formation. Models incorporating these processes are very demanding and are under active development. 

\subsubsection{1D spherically-symmetric simulations}
To reduce the complexity of modelling \ac{CE} evolution, the first detailed simulations were performed assuming the donor star's envelope is spherically symmetric \citep{meyer1979a,taam1978a,taam1979a}. With Eq.~(\ref{eq:Fdrag}) or more elaborate expressions for the drag force \citep[e.g.,][]{bondi1952a,ostriker1999a,kim2007a}, it is possible to formulate a model for the inspiral through a 1D \ac{CE}\footnote{In such models, the envelope is treated as a spherically-symmetric (1D) structure, but the path of the inspiral can involve radial and lateral components and therefore Eq.~(\ref{eq:CE_1D}) is given in the general vector form.} from the equations of motion for the donor core and the companion \cite[e.g.,][]{bronner2024a},
\begin{align}
  \ddot{\mathbf{r}} = - \frac{G \left(M_{1,r} + M_2\right)}{r^2} \frac{\mathbf{r}}{r} - \frac{F_\mathrm{drag}}{M_2} \frac{\dot{\mathbf{r}}}{\dot{r}},
  \label{eq:CE_1D}
\end{align}
where $\mathbf{r}=\mathbf{r}_2-\mathbf{r}_1$ is the separation between the companion and the core of the donor, $M_{1,r}$ is the donor's mass inside the radius $r$, and $F_\mathrm{drag}$ is the drag force acting on the companion. The orbital energy that is dissipated by the drag force can be added as heat into the envelope at the companion's position.

Compared to more elaborate simulations (in three spatial dimensions as discussed in Section~\ref{sect:CE_3D}), such 1D spherically-symmetric models of \ac{CE} evolution are substantially less demanding in terms of the required computational resources. This makes them an interesting test bench for physical effects that may otherwise be computationally expensive to incorporate, such as energy transport, nuclear burning, and the response of the core of the donor to envelope mass loss. With 1D models, it is also possible to follow \ac{CE} evolution for much longer timescales after the plunge-in. For example, the transition of \ac{CE} evolution towards a \textit{self-regulated inspiral}, where the orbital energy released by the binary balances radiative losses at the photosphere, has been simulated in 1D \citep{meyer1979a,clayton2017a}. Finally, the outcome of 1D simulations can be directly mapped back to classical, 1D stellar evolution codes to follow the subsequent nuclear-timescale evolution of the post-\ac{CE} binary.

Nonetheless, there are gross simplifications made by 1D spherically-symmetric \ac{CE} models. For instance, spherical symmetry implies energy is deposited into spherical shells. It is also not possible to capture the drag force acting on the donor's core. Moreover, the available (semi-)analytic expressions for $F_\mathrm{drag}$ usually provide only a scaling with characteristic quantities of the system rather than an exact value. They have to be calibrated with free multiplicative parameters, which limits the predictive power of 1D inspiral models for \ac{CE} evolution.

\subsubsection{3D simulations}
\label{sect:CE_3D}

\begin{figure}[tb]
    \centering
    \includegraphics[width=1.0\textwidth]{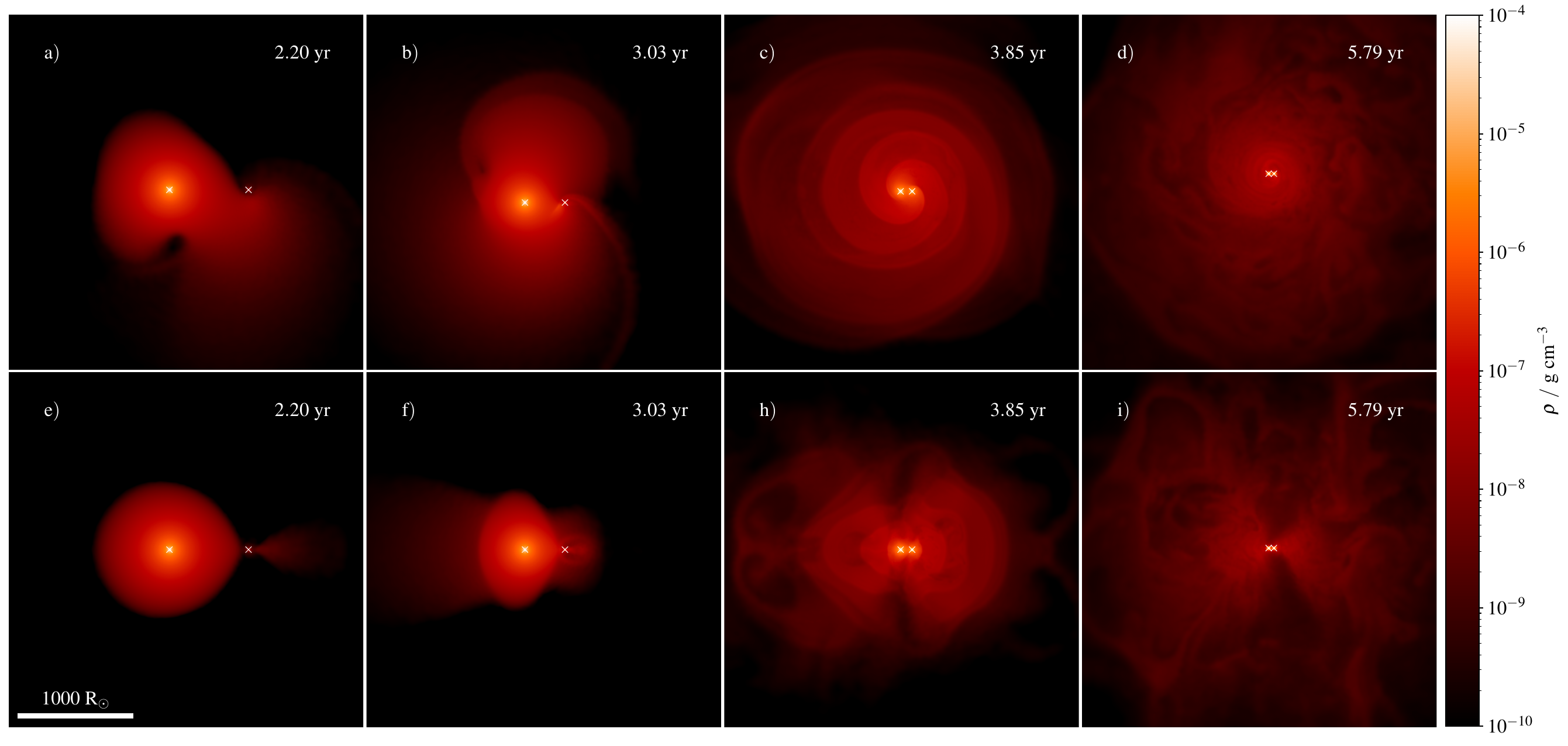}
    \caption{A 3D hydrodynamical simulation of \ac{CE} evolution, showing the evolution of density, $\rho$, in the orbital plane (top panel) and from an edge-on view of the plane containing the stellar cores (bottom plane). The donor is a 12\msun red supergiant while the 3\msun companion is represented as a point mass. The donor core and the companion are shown as cross markers. This plot is based on the simulations of \cite{lau2022a} and was made using the \texttt{Sarracen} python library \citep{harris2023a}.}
    \label{fig:3d_CE}
\end{figure}

The shortcomings of 1D spherically-symmetric \ac{CE} models can be overcome with detailed 3D hydrodynamic simulations. Various types of \ac{CE} interactions have been simulated in 3D, including high- and low-mass primary stars at different evolutionary stages with companions that could represent \ac{MS} stars, white and brown dwarf stars, neutron stars and black holes \citep[e.g.,][]{passy2012a,ricker2012a,nandez2014a,ohlmann2016a,prust2019a,chamandy2018a,kramer2020a,sand2020a,ondratschek2022a,lau2022a,moreno2022a}. \ac{CE} interaction in triple systems has been simulated \citep{glanz2021b}, as well as the related process of planetary engulfment that may occur when a star expands during its giant phase and captures its innermost planets \citep{staff2016a,chamandy2018a,lau2022c}.

An example for a 3D \ac{CE} simulation is shown in Fig.~\ref{fig:3d_CE}. At first, the companion significantly distorts the donor as it approaches its surface (Fig. \ref{fig:3d_CE}a), launching unbound ejecta in the wake of the companion. After entering the envelope, the companion rapidly plunges inwards. The orbital motion of the companion and donor core around their centre of mass inside the \ac{CE} is supersonic, launching spiral shock waves in the orbital plane (Fig. \ref{fig:3d_CE}c). In the further evolution, this spiral structure winds up and tightens due to increasing orbital frequency. The inspiral of the two cores eventually slows down and the orbital separation settles to a value that changes only very slowly compared to the orbital period ($a/(-\dot{a}) \gg P_\mathrm{orb}$). The stalling of the orbital separation is associated with a lowered density of envelope gas in the vicinity and a greater degree of co-rotation with this gas, which both contribute to reducing gravitational drag. The shape of the envelope significantly diverges from spherical symmetry and assumes a toroidal shape (Fig. \ref{fig:3d_CE}h). Shear instabilities emerge and overlay the spiral pattern (Fig. \ref{fig:3d_CE}d), and are observed to disrupt the spiral morphology produced during the plunge-in. In the case of successful envelope ejection, the density of the material in the toroidal structure decreases as the material expands and becomes gravitationally unbound from the system.

\subsection{Success and failure of envelope ejection}
A common finding of 3D simulations of CE evolution based on a model only accounting for hydrodynamics and gravity is that the system usually fails to eject the envelope. Additional processes must be taken into account to achieve full envelope ejection. One of such processes is the release of ionisation energy in the envelope material when it expands and its ions recombine with free electrons. If the released recombination energy thermalises inside the \ac{CE}, it builds up a pressure gradient that accelerates the material outwards and supports envelope ejection. Loss of all envelope material is possible in such cases, but it happens significantly after the settling of the orbital separation to its final value.

Even in cases where the envelope remains bound, it is lifted from the stellar cores and establishes  a toroidal structure around them. This reduces the gravitational drag acting on the cores and the orbit changes only very slowly after rapid plunge-in. The still-bound envelope material, however, could form a persistent circumbinary disk that continues to interact with the central binary. Eventually, this could lead to a merger of the two stellar cores \citep{wei2024a}.

\subsection{Magnetic field amplification}\label{chap1:ce-bfields}

\begin{figure}[tb]
    \centering
    \includegraphics[width=1.0\textwidth]{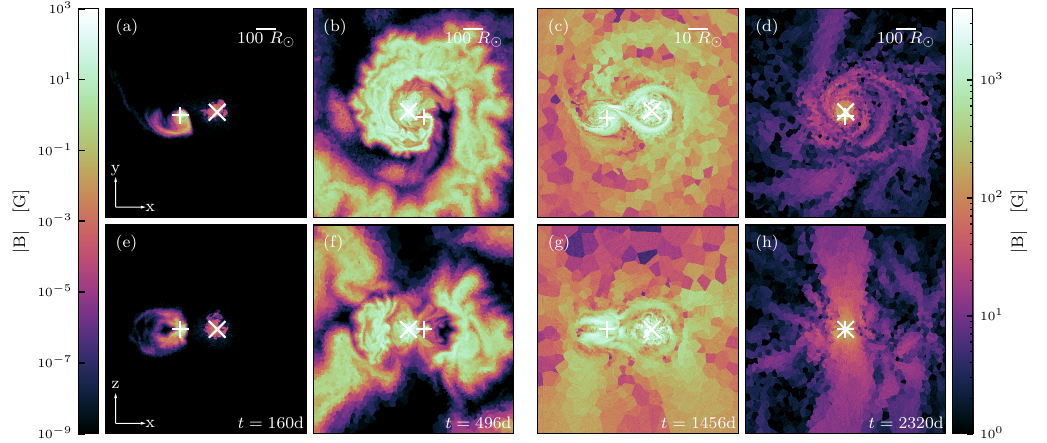}
    \caption{Evolution of the absolute magnetic field strength of the magnetohydrodynamic \ac{CE} simulation performed by \citet{ondratschek2022a} in the face-on (top) and edge-on (bottom) view. Times of the snapshot and spatial scales are given in the respective panels. The core of the primary star is marked with a ``$\times$'' and the companion is marked with a ``$+$''. Image courtesy of Marco Vetter.}
    \label{fig:mhd-ce-bfield}
\end{figure}

Similar to stellar mergers, magnetohydrodynamic simulations of \ac{CE} evolution find that strong amplification of initial magnetic fields can take place. In the first such simulation, \citet{ohlmann2016b} point out the relevance of the magnetorotational instability for amplifying a weak initial magnetic field by many orders of magnitude. This process is shown in Fig.~\ref{fig:mhd-ce-bfield}. In contrast to the situation observed in stellar mergers (see Sect.~\ref{chap1:mergers:b-fields}), the field amplification at the beginning of the \ac{CE} plunge-in is a localised process around the stellar cores (Fig.~\ref{fig:mhd-ce-bfield}(a),(e)). After several orbits (Fig.~\ref{fig:mhd-ce-bfield}(b),(f)), the field amplification proceeds along the spiral shock waves that permeate the entire \ac{CE}. After the end of the plunge-in, there is a transfer of material that has accumulated in the vicinity of the stellar cores, and the resulting shear amplifies the magnetic fields in a way similar to the case of stellar mergers (Fig.~\ref{fig:mhd-ce-bfield}(c),(g)). As with the case of stellar mergers, the magnetic field strength saturates towards the end of the plunge-in.

\begin{figure}[tb]
    \centering
    \includegraphics[width=0.7\textwidth]{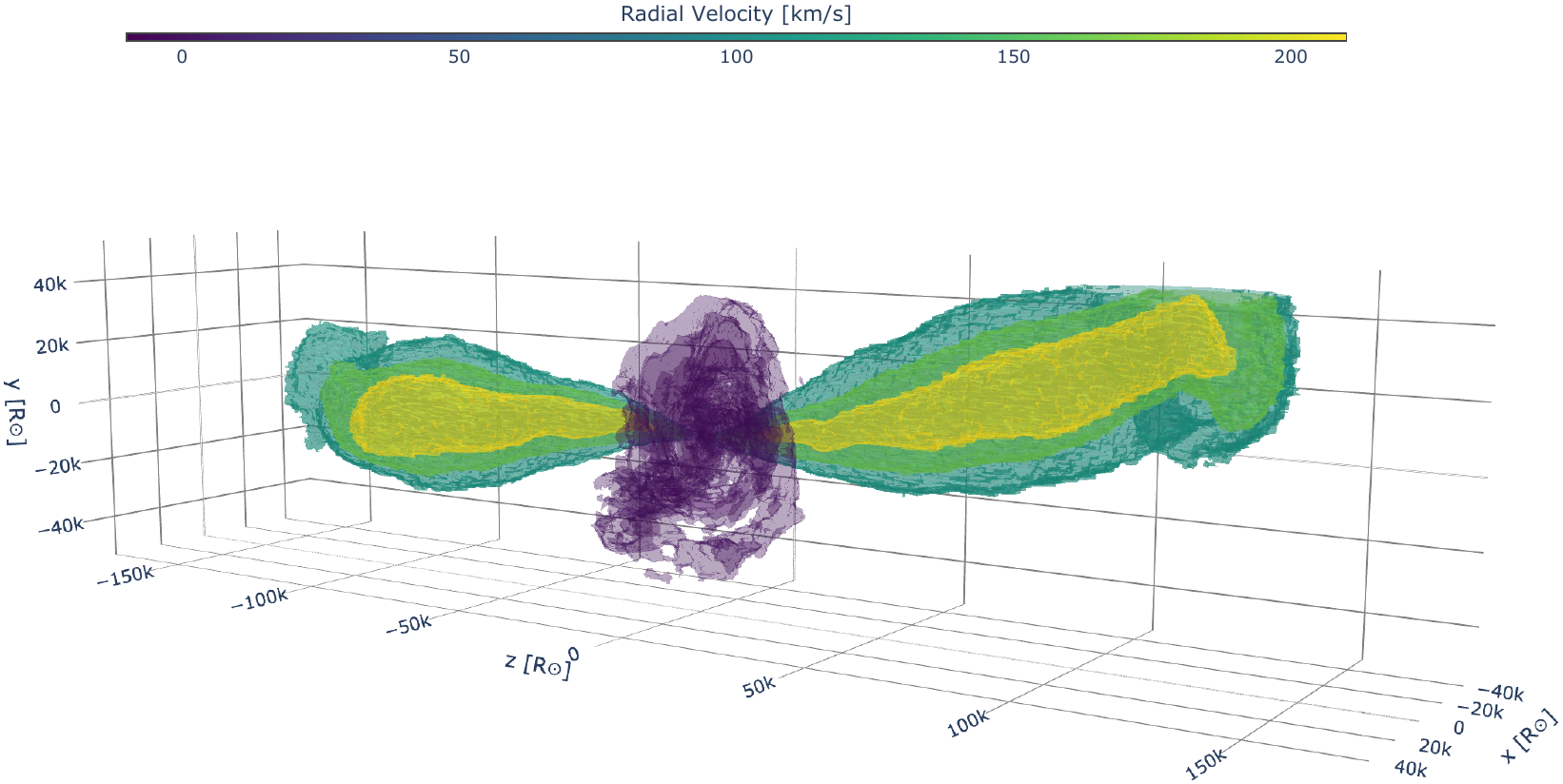}
    \caption{3D rendering of the post-plunge-in structure obtained in a magnetohydrodynamic simulation of the \ac{CE} interaction of a $5\,\mathrm{M}_\odot$ black hole with a $10\,\mathrm{M}_\odot$ red supergiant \citep[from the simulation by][]{moreno2022a, vetter2024a}. Color-coded is the radial velocity component with yellow/green/blue colors indicating outward and purple colors indicating inward flows. Image courtesy of Marco Vetter.}
    \label{fig:jet}
\end{figure}

\citet{ohlmann2016b} find that magnetic fields have a limited effect on the orbital evolution and amount of ejected envelope mass for the system they simulated. Despite this, they still produce a significant dynamical effect that imprints on the ejecta morphology. \citet{ondratschek2022a} performed magnetohydrodynamic simulations of \ac{CE}  evolution with an asymptotic giant branch primary star and find high-velocity bipolar outflows that are not observed in a purely hydrodynamic simulation. These jet-like outflows set in after the dynamical plunge-in, when the binary semi-major axis has stopped decreasing significantly. \citet{ondratschek2022a} attribute this phenomenon to magnetic launching of material through a cavity in the toroidal envelope structure as also seen in panel (h) of Fig.~\ref{fig:mhd-ce-bfield} and in Fig.~\ref{fig:jet}. In this jet, through which a fraction of the envelope is expelled, material is highly magnetised and is ejected along wound helical field lines with velocities exceeding some $100 \, \mathrm{km}\, \mathrm{s}^{-1}$. This is much larger than the ejection velocities of the majority of the \ac{CE} material, which are of the order of $10 \, \mathrm{km}\, \mathrm{s}^{-1}$). Some simulations \citep{vetter2024a} show that in this stage, the morphology of the system transitions towards a disk-jet structure, in which a thick magnetised accretion disk around the central binary feeds a jet-like outflow (see Fig.~\ref{fig:jet}). Such a structure is consistent with the bipolar morphology observed in some planetary nebulae (see Section \ref{subsec:post-CE-binaries}) and ``water fountains'' in young pre-planetary nebulae. The morphology of planetary nebulae formed from ejected \acp{CE} would be further shaped by stellar winds launched by the remnant core over timescales of $\sim10^4\, \mathrm{yr}$. The magnetic field amplification, jet launching, and a restructuring of the system still need to be investigated in more detail in simulations, but it appears likely that strongly amplified magnetic fields are important for shaping the morphology of post-CE systems.

\subsection{Shortcomings and challenges of current common-envelope simulations}

Current 3D simulations of \ac{CE} interaction as described above suffer from several shortcomings in modelling the physical processes at work. One of these concerns the effect of recombination energy release. The assumption of local thermalisation of this energy cannot be justified at all epochs of \ac{CE} evolution and throughout the envelope material. Dilute regions become transparent to radiation that can efficiently carry away energy without supporting envelope unbinding. Models for radiation transport are being implemented in \ac{CE} simulations, but they face challenges. The transition between the optically thin and thick regions in the envelope material (the photosphere) is difficult to resolve spatially in numerical representations. Moreover, it is suspected to be very close to the hydrogen recombination front, the region where hydrogen ions recombine.

Another shortcoming is that current 3D simulations are usually performed with unrealistic initial conditions with the companion already very near or touching the surface of the donor star. This means the binary promptly enters a dynamical spiral-in, but the preceding prolonged onset of dynamically unstable mass transfer is not captured. 

With 3D simulations, it is difficult to follow \ac{CE} evolution on long timescales. In the first place, this is related to the strict timestep restriction for numerical stability in hydrodynamic simulations, which only allows following the interaction on dynamical timescales, but not on much longer timescales which may govern the post-plunge-in evolution of the systems. In this phase of the evolution, additional physical processes need to be taken into account, such as radiation transfer and dust formation \citep{glanz2018a,bermudez-bustamante2024a}. Moreover, with many time integration steps taken, a substantial buildup of energy or angular momentum errors is observed.

The high computational costs of detailed 3D simulations make resolution studies challenging. Numerical convergence of key quantities like the amount of unbound envelope mass and final orbital separation have yet to be demonstrated. This raises the question of whether the observed physical phenomena are represented in a reliable way. Convective transport of energy is a particular concern in this context.

Perhaps the most fundamental drawback of 3D simulations is that the core of the donor and the companion are usually represented as point particles that only interact via gravity to remedy the spatial scale challenge. This prevents a realistic model of the reaction of the cores to the envelope material and its removal. Effects such as accretion or the expansion of outer layers of the core once the envelope is lifted are currently not consistently treated in 3D simulations. The point particles must also interact via a \emph{softened} gravity law, meaning they have artificial smooth potential wells instead of following the Newtonian inverse-square law, which would cause gravity to diverge at the centre. This procedure limits the extent to which material could be compressed around a point particle, which then behaves as an object with finite size. The characteristic radius of such softened gravitational wells is typically the smallest resolution scale and therefore sets the largest time step that can be taken in a \ac{CE} simulation. A tractable calculation requires the softening radius to greatly exceed, e.g., the helium core of a red giant or the radius of an in-spiralling neutron star. This prevents simulations from resolving the response of the core to mass loss or processes occurring near it, such as accretion onto a neutron star or black hole companion.

\subsection{Luminous red novae}
\label{subsec:LRNe}

\acp{CE} and stellar mergers have been proposed to give rise to a class of electromagnetic transients broadly known as \emph{\acp{LRN}}\footnote{\cite{pastorello2019b} distinguishes between a \ac{LRN} and a red nova. The latter shares many common characteristics with a \ac{LRN} but is less luminous, and so is likely produced by mergers of lower-mass binaries as in V1309 Sco. We broadly refer to both types of transient as \acp{LRN}.}. To date, more than a dozen candidates for \acp{LRN} have been observed and characterised. They broadly share the following characteristics: (i) A week- to month-long red plateau that evolves towards the infrared domain, (ii) A large range of plateau luminosities with a range of $\sim 10^{36}-10^{42}~\mathrm{erg}~\mathrm{s}^{-1}$ ($-3>M_V>-16~\mathrm{mag}$), and (iii) A shorter initial blue peak.

The association of \acp{LRN} with \acp{CE} and steller mergers had only been largely speculative for a long time. This connection is at least consistent from an energetics perspective. If the amount of energy that can be liberated and converted into radiation is of the order of the orbital energy\footnote{This assumption is agnostic to the specific emission mechanism, which is actively researched and required to obtain accurate lightcurve models.}, $E_\mathrm{orb}\sim GM^2/R$ at the characteristic stellar radius $R$, and the energy is radiated on the dynamical timescale, $P_\mathrm{orb}\sim [R^3/(GM)]^{1/2}$, then the available luminosity is
\begin{align}
    \frac{E_\mathrm{orb}}{P_\mathrm{orb}}
    \sim \frac{G^{3/2}M^{5/2}}{R^{5/2}}
    \sim 10^{40}~\bigg(\frac{M}{\mathrm{M}_\odot}\bigg)^{5/2} \bigg(\frac{R}{100~\mathrm{R}_\odot}\bigg)^{-5/2}~\mathrm{erg}~\mathrm{s}^{-1},
    \label{eq:LRN}
\end{align}
consistent with the plateau luminosities of more energetic \acp{LRN}. Calculating the actual observed luminosity is more complicated as it depends on the evolution of the luminosity and size of the emitting region. While a radius of $100~\mathrm{R}_\odot$ has been adopted in Eq.~(\ref{eq:LRN}), variations in the stellar masses and compactness can give rise to the range of observed luminosities for this class of transients. Recombination energy released by the ejected material may also power the luminosity of a \ac{LRN} \citep{ivanova2013b,matsumoto2022a}, in which case Eq.~(\ref{eq:LRN}) is not directly applicable.

A confirmation that stellar mergers can lead to \acp{LRN} was only provided by the landmark outburst of V1309 Sco. V1309 Sco was a Galactic contact binary with a 1.4-day orbital period that was monitored with high-cadence photometry under the OGLE project from 2001 \citep{tylenda2011a}. Its lightcurve suggested an exponential period decrease over the next seven years, presumably merging in 2008 when it produced a red nova with peak absolute magnitude of $M_V=-6.56$ and a bolometric luminosity of $\approx 2\times10^4\lsun$ \citep{nakano2008a,mason2010b}. V1309 Sco is now thought to have been composed of a slightly evolved, approximately $1.5\msun$ star and a much less massive \ac{MS} star, and might have been brought to merge by the Darwin instability. The slightly evolved donor star is unlikely to have a clearly distinguished core and envelope, therefore aligning with a post-\ac{MS} merger rather than a \ac{CE} event, following the discussion in Sections \ref{chap1:introduction} and \ref{chap1:mergers:ms+ms-vs-postMS+MS-mergers}. This remarkably well observed event has been modelled in many works, which aim to discover the physical processes like $L_2$ outflows, recombination, shock-heating, and dust formation that give rise to the observed features. Another crucial \ac{LRN} is the outburst of V838 Mon in 2002 \citep{munari2002a,bond2003a}, which shares many similar characteristics with V1309 Sco and is believed to be a merger involving a more massive primary star of $\sim 8\msun$. The light echo of V838 Mon, produced from the light reflected on surrounding interstellar dust, has been famously captured in a series of images taken by the Hubble Space Telescope. Many more \acp{LRN}, up to a few hundred per year \citep{howitt2020a}, are expected to be observed in the near future by the Vera C. Rubin Observatory.

\subsection{``Post-common-envelope'' binaries}
\label{subsec:post-CE-binaries}

While a great variety of compact binaries are likely to have formed from \ac{CE} evolution, systems that have emerged from a \ac{CE} with minimal further evolution provide valuable constraints on the outcome of \ac{CE} evolution. Such is the case for V471 Tau, and, today, many more such short-period detached binaries suspected to have gone through a \ac{CE} have been found. They are composed of a \ac{MS} star and a white dwarf or O/B-type subdwarf (sdO/B) star, and are commonly referred to as \emph{post-\ac{CE} binaries}. These observations have been used to test the energy formalism (Eq. \ref{eq:alpha}) and constrain the uncertain \ac{CE} efficiency parameter, $\alpha_\mathrm{CE}$ \citep{zorotovic2010a,demarco2011a}. In the expression for the change in orbital energy (Eq. \ref{eq:Eorb}), $M_\mathrm{1,core}$ is taken to be the white dwarf or sdO/B mass, $M_2$ is the mass of the \ac{MS} star, and $a_f$ is the present-day binary separation. Inferring the value for $\alpha_\mathrm{CE}$ still requires knowing $a_i$, which is the radius of a giant star, and the envelope binding energy $E_\mathrm{bind}$. These may be determined using stellar evolution calculations to find a red giant model that has a core mass equal to the white dwarf mass. This is possible due to the tight relationship between giant radii/luminosities and core masses \citep[e.g.,][]{joss1987a}.

The great majority of post-\ac{CE} binaries are found to give $\alpha_\mathrm{CE}$ values of approximately unity, though many 3D \ac{CE} simulations find final separations that are significantly larger. There are significant uncertainties from such inferences \citep{iaconi2019a}. One source of error is that the post-\ac{CE} binaries could have been further tightened through gravitational radiation and magnetic braking, which would falsely shift the inferred $\alpha_\mathrm{CE}$ value downwards. However, post-\ac{CE} binaries have also been found around planetary nebulae, which are interpreted as \acp{CE} ejected no longer than $\sim 10^4~\mathrm{yr}$ in the past before the nebula can be dispersed. Possibly $\approx 20\%$ or more planetary nebulae could be ejected \acp{CE}, based on the close binary fraction measured from photometric variability surveys \citep{miszalski2009a,jacoby2021a}. Diverse morphology of these planetary nebulae has been found, including elongated and bipolar morphologies suggesting the presence of bipolar outflows \citep{demarco2009a}.

\section{Open questions and future directions}\label{chap1:open-questions}

The dynamic interaction of two stars in a binary system via mergers and CE phases still leaves many open questions. While the basic physical picture is becoming clearer, several physical processes still need to be studied in greater detail and ongoing research with detailed simulations tries to tackle them. For example, radiation is relevant for energy transport and photon cooling in massive star mergers and virtually in all CE events. Similarly, magnetic fields are amplified to such values that they are found to drive fast, bipolar outflows (jets), but the impact of such outflows on the merger and CE dynamics is not yet well understood. Moreover, convective fluid motions may be relevant (particularly in CE events) and circumstellar/circumbinary discs may form around the merger remnant and an inner post-CE binary. The interaction of such discs with the immediate merger remnant or inner post-CE binary is not yet fully understood but has the potential to significantly influence the outcome of such dynamic binary interactions.

With an ever more complete understanding of stellar mergers and CE events, it is possible to develop better 1D models for inclusion in binary star evolution computations and ultimately better prescriptions for binary population synthesis simulations. Taking such steps is key to connecting to observed populations of merged stars and post-CE systems. However, going from 3D to 1D models is challenging and may be physically impossible as much of the physics in stellar mergers and CE events are inherently 3D processes. Still, simplified and effective models that can reproduce certain key outcomes are already very insightful and will further the field. Connecting 3D models with 1D binary star evolution is not a one-way street. The initial conditions of the 3D simulations come directly from binary evolution computations and, as detailed in Sect.~\ref{chap1:contact-binaries}, the question of which binary stars end up in contact phases is crucial and linked to, \eg, the still uncertain stability of binary mass transfer. 

Another important future step is to relate stellar mergers and CE phases to the transients observed as luminous red novae. At the moment, this is very difficult as most 3D simulations of stellar mergers and CE events do not properly include radiation transport and thus can not accurately predict the electromagnetic transient signal very accurately. One of the biggest challenges will be to properly model the transition from optically thick regions to the optically thin photosphere. Such studies will be complicated, \eg, by asymmetric structures, complex cooling processes, and dust physics.

\begin{ack}[Acknowledgments]
~The authors acknowledge support by the Klaus Tschira Foundation. This work has received funding from the European Research Council (ERC) under the European Union's Horizon 2020 research and innovation programme (Grant agreement No.\ 945806) and by the European Union (ERC, ExCEED, project number 101096243). Views and opinions expressed are, however, those of the authors only and do not necessarily reflect those of the European Union or the European Research Council Executive Agency. Neither the European Union nor the granting authority can be held responsible for them. This work was supported by the Deutsche Forschungsgemeinschaft (DFG, German Research Foundation) under Germany's Excellence Strategy EXC 2181/1-390900948 (the Heidelberg STRUCTURES Excellence Cluster). M. Y. M. L. is supported by a Croucher Fellowship awarded by the Croucher Foundation.
\end{ack}

\seealso{\citet{langer2012a}, \citet{ivanova2013a}, \citet{ropke2023a}, \citet{marchant2024a}}

\bibliographystyle{Harvard}
\begin{thebibliography*}{126}
    \providecommand{\bibtype}[1]{}
    \providecommand{\natexlab}[1]{#1}
    {\catcode`\|=0\catcode`\#=12\catcode`\@=11\catcode`\\=12
    |immediate|write|@auxout{\expandafter\ifx\csname
      natexlab\endcsname\relax\gdef\natexlab#1{#1}\fi}}
    \renewcommand{\url}[1]{{\tt #1}}
    \providecommand{\urlprefix}{URL }
    \expandafter\ifx\csname urlstyle\endcsname\relax
      \providecommand{\doi}[1]{doi:\discretionary{}{}{}#1}\else
      \providecommand{\doi}{doi:\discretionary{}{}{}\begingroup
      \urlstyle{rm}\Url}\fi
    \providecommand{\bibinfo}[2]{#2}
    \providecommand{\eprint}[2][]{\url{#2}}
    
    \bibtype{Article}%
    \bibitem[Balbus(1995)]{balbus1995a}
    \bibinfo{author}{Balbus SA} (\bibinfo{year}{1995}), \bibinfo{month}{Nov.}
    \bibinfo{title}{General local stability criteria for stratified, weakly
      magnetized rotating systems}.
    \bibinfo{journal}{{\em Astrophysical Journal}} \bibinfo{volume}{453}:
      \bibinfo{pages}{380}.
    
    \bibtype{Article}%
    \bibitem[Balbus and Hawley(1991)]{balbus1991a}
    \bibinfo{author}{Balbus SA} and  \bibinfo{author}{Hawley JF}
      (\bibinfo{year}{1991}), \bibinfo{month}{Jul.}
    \bibinfo{title}{A powerful local shear instability in weakly magnetized disks.
      {I} - {Linear} analysis. {II} - {Nonlinear} evolution}.
    \bibinfo{journal}{{\em Astrophysical Journal}} \bibinfo{volume}{376}:
      \bibinfo{pages}{214--233}.
    
    \bibtype{Article}%
    \bibitem[Bellinger et al.(2024)]{bellinger2024a}
    \bibinfo{author}{Bellinger EP}, \bibinfo{author}{de~Mink SE},
      \bibinfo{author}{van Rossem WE} and  \bibinfo{author}{Justham S}
      (\bibinfo{year}{2024}), \bibinfo{month}{Jun.}
    \bibinfo{title}{The {Potential} of {Asteroseismology} to {Resolve} the {Blue}
      {Supergiant} {Problem}}.
    \bibinfo{journal}{{\em The Astrophysical Journal}} \bibinfo{volume}{967}:
      \bibinfo{pages}{L39}.
    
    \bibtype{Article}%
    \bibitem[Bermúdez-Bustamante et al.(2024)]{bermudez-bustamante2024a}
    \bibinfo{author}{Bermúdez-Bustamante LC}, \bibinfo{author}{De~Marco O},
      \bibinfo{author}{Siess L}, \bibinfo{author}{Price DJ},
      \bibinfo{author}{González-Bolívar M}, \bibinfo{author}{Lau MYM},
      \bibinfo{author}{Mu C}, \bibinfo{author}{Hirai R},
      \bibinfo{author}{Danilovich T} and  \bibinfo{author}{Kasliwal MM}
      (\bibinfo{year}{2024}), \bibinfo{month}{Sep.}
    \bibinfo{title}{Dust formation in common envelope binary interactions - {II}:
      {3D} simulations with self-consistent dust formation}.
    \bibinfo{journal}{{\em Monthly Notices of the Royal Astronomical Society}}
      \bibinfo{volume}{533}: \bibinfo{pages}{464--481}.
    
    \bibtype{Article}%
    \bibitem[Bernini-Peron et al.(2023)]{bernini-peron2023a}
    \bibinfo{author}{Bernini-Peron M}, \bibinfo{author}{Marcolino WLF},
      \bibinfo{author}{Sander AAC}, \bibinfo{author}{Bouret JC},
      \bibinfo{author}{Ramachandran V}, \bibinfo{author}{Saling J},
      \bibinfo{author}{Schneider FRN}, \bibinfo{author}{Oskinova LM} and
      \bibinfo{author}{Najarro F} (\bibinfo{year}{2023}), \bibinfo{month}{Sep.}
    \bibinfo{title}{Clumping and {X}-rays in cooler {B} supergiant stars}.
    \bibinfo{journal}{{\em Astronomy and Astrophysics}} \bibinfo{volume}{677}:
      \bibinfo{pages}{A50}.
    
    \bibtype{Article}%
    \bibitem[Bond et al.(2003)]{bond2003a}
    \bibinfo{author}{Bond HE}, \bibinfo{author}{Henden A}, \bibinfo{author}{Levay
      ZG}, \bibinfo{author}{Panagia N}, \bibinfo{author}{Sparks WB},
      \bibinfo{author}{Starrfield S}, \bibinfo{author}{Wagner RM},
      \bibinfo{author}{Corradi RLM} and  \bibinfo{author}{Munari U}
      (\bibinfo{year}{2003}), \bibinfo{month}{Mar.}
    \bibinfo{title}{An energetic stellar outburst accompanied by circumstellar
      light echoes}.
    \bibinfo{journal}{{\em Nature}} \bibinfo{volume}{422}:
      \bibinfo{pages}{405--408}.
    
    \bibtype{Article}%
    \bibitem[Bondi(1952)]{bondi1952a}
    \bibinfo{author}{Bondi H} (\bibinfo{year}{1952}), \bibinfo{month}{Jan.}
    \bibinfo{title}{On spherically symmetrical accretion}.
    \bibinfo{journal}{{\em Monthly Notices of the Royal Astronomical Society}}
      \bibinfo{volume}{112}: \bibinfo{pages}{195}.
    
    \bibtype{Article}%
    \bibitem[Braun and Langer(1995)]{braun1995a}
    \bibinfo{author}{Braun H} and  \bibinfo{author}{Langer N}
      (\bibinfo{year}{1995}), \bibinfo{month}{May}.
    \bibinfo{title}{Effects of accretion onto massive main sequence stars.}
    \bibinfo{journal}{{\em Astronomy and Astrophysics}}
    
    \bibtype{Article}%
    \bibitem[Britavskiy et al.(2019)]{britavskiy2019a}
    \bibinfo{author}{Britavskiy N}, \bibinfo{author}{Lennon DJ},
      \bibinfo{author}{Patrick LR}, \bibinfo{author}{Evans CJ},
      \bibinfo{author}{Herrero A}, \bibinfo{author}{Langer N}, \bibinfo{author}{van
      Loon JT}, \bibinfo{author}{Clark JS}, \bibinfo{author}{Schneider FRN} and
      \bibinfo{author}{Almeida LA} (\bibinfo{year}{2019}), \bibinfo{month}{Apr.}
    \bibinfo{title}{The {VLT}-{FLAMES} tarantula survey. {XXX}. {Red} stragglers in
      the clusters hodge 301 and {SL} 639}.
    \bibinfo{journal}{{\em Astronomy and Astrophysics}} \bibinfo{volume}{624}:
      \bibinfo{pages}{A128}.
    
    \bibtype{Article}%
    \bibitem[Bronner et al.(2024)]{bronner2024a}
    \bibinfo{author}{Bronner VA}, \bibinfo{author}{Schneider FRN},
      \bibinfo{author}{Podsiadlowski P} and  \bibinfo{author}{Röpke FK}
      (\bibinfo{year}{2024}), \bibinfo{month}{Mar.}
    \bibinfo{title}{Going from {3D} to {1D}: {A} one-dimensional approach to
      common-envelope evolution}.
    \bibinfo{journal}{{\em Astronomy and Astrophysics}} \bibinfo{volume}{683}:
      \bibinfo{pages}{A65}.
    
    \bibtype{Article}%
    \bibitem[Chamandy et al.(2018)]{chamandy2018a}
    \bibinfo{author}{Chamandy L}, \bibinfo{author}{Frank A},
      \bibinfo{author}{Blackman EG}, \bibinfo{author}{Carroll-Nellenback J},
      \bibinfo{author}{Liu B}, \bibinfo{author}{Tu Y}, \bibinfo{author}{Nordhaus
      J}, \bibinfo{author}{Chen Z} and  \bibinfo{author}{Peng B}
      (\bibinfo{year}{2018}), \bibinfo{month}{Oct.}
    \bibinfo{title}{Accretion in common envelope evolution}.
    \bibinfo{journal}{{\em Monthly Notices of the Royal Astronomical Society}}
      \bibinfo{volume}{480} (\bibinfo{number}{2}): \bibinfo{pages}{1898--1911}.
    
    \bibtype{Article}%
    \bibitem[Chandrasekhar(1943)]{chandrasekhar1943a}
    \bibinfo{author}{Chandrasekhar S} (\bibinfo{year}{1943}), \bibinfo{month}{Mar.}
    \bibinfo{title}{Dynamical friction. {I}. {General} considerations: the
      coefficient of dynamical friction.}
    \bibinfo{journal}{{\em Astrophysical Journal}} \bibinfo{volume}{97}:
      \bibinfo{pages}{255}.
    
    \bibtype{Article}%
    \bibitem[Claeys et al.(2011)]{claeys2011a}
    \bibinfo{author}{Claeys JSW}, \bibinfo{author}{de~Mink SE},
      \bibinfo{author}{Pols OR}, \bibinfo{author}{Eldridge JJ} and
      \bibinfo{author}{Baes M} (\bibinfo{year}{2011}), \bibinfo{month}{Apr.}
    \bibinfo{title}{Binary progenitor models of type {IIb} supernovae}.
    \bibinfo{journal}{{\em Astronomy and Astrophysics}} \bibinfo{volume}{528}:
      \bibinfo{pages}{A131}.
    
    \bibtype{Article}%
    \bibitem[Clayton et al.(2017)]{clayton2017a}
    \bibinfo{author}{Clayton M}, \bibinfo{author}{Podsiadlowski P},
      \bibinfo{author}{Ivanova N} and  \bibinfo{author}{Justham S}
      (\bibinfo{year}{2017}), \bibinfo{month}{Sep.}
    \bibinfo{title}{Episodic mass ejections from common-envelope objects}.
    \bibinfo{journal}{{\em Monthly Notices of the Royal Astronomical Society}}
      \bibinfo{volume}{470} (\bibinfo{number}{2}): \bibinfo{pages}{1788--1808}.
    
    \bibtype{Article}%
    \bibitem[Colgate(1967)]{colgate1967a}
    \bibinfo{author}{Colgate SA} (\bibinfo{year}{1967}), \bibinfo{month}{Oct.}
    \bibinfo{title}{Stellar coalescence and the multiple supernova interpretation
      of quasi-stellar sources}.
    \bibinfo{journal}{{\em Astrophysical Journal}} \bibinfo{volume}{150}:
      \bibinfo{pages}{163}.
    
    \bibtype{Article}%
    \bibitem[Costa et al.(2022)]{costa2022a}
    \bibinfo{author}{Costa G}, \bibinfo{author}{Ballone A},
      \bibinfo{author}{Mapelli M} and  \bibinfo{author}{Bressan A}
      (\bibinfo{year}{2022}), \bibinfo{month}{Oct.}
    \bibinfo{title}{Formation of black holes in the pair-instability mass gap:
      {Evolution} of a post-collision star}.
    \bibinfo{journal}{{\em Monthly Notices of the Royal Astronomical Society}}
      \bibinfo{volume}{516} (\bibinfo{number}{1}): \bibinfo{pages}{1072--1080}.
    
    \bibtype{Article}%
    \bibitem[Darwin(1879)]{darwin1879a}
    \bibinfo{author}{Darwin GH} (\bibinfo{year}{1879}).
    \bibinfo{title}{The determination of the secular effects of tidal friction by a
      graphical method}.
    \bibinfo{journal}{{\em Proceedings of the Royal Society of London}}
      \bibinfo{volume}{29} (\bibinfo{number}{196-199}): \bibinfo{pages}{168--181}.
    
    \bibtype{Article}%
    \bibitem[De~Marco(2009)]{demarco2009a}
    \bibinfo{author}{De~Marco O} (\bibinfo{year}{2009}), \bibinfo{month}{Apr.}
    \bibinfo{title}{The origin and shaping of planetary nebulae: {Putting} the
      binary hypothesis to the test}.
    \bibinfo{journal}{{\em Publications of the ASP}} \bibinfo{volume}{121}:
      \bibinfo{pages}{316--342}.
    
    \bibtype{Article}%
    \bibitem[De~Marco et al.(2011)]{demarco2011a}
    \bibinfo{author}{De~Marco O}, \bibinfo{author}{Passy JC}, \bibinfo{author}{Moe
      M}, \bibinfo{author}{Herwig F}, \bibinfo{author}{Mac~Low MM} and
      \bibinfo{author}{Paxton B} (\bibinfo{year}{2011}), \bibinfo{month}{Mar.}
    \bibinfo{title}{On the {\textless}span
      class="nocase"{\textgreater}α{\textless}/span{\textgreater} formalism for
      the common envelope interaction}.
    \bibinfo{journal}{{\em Monthly Notices of the Royal Astronomical Society}}
      \bibinfo{volume}{411}: \bibinfo{pages}{2277--2292}.
    
    \bibtype{Article}%
    \bibitem[De~Young(1968)]{deyoung1968a}
    \bibinfo{author}{De~Young DS} (\bibinfo{year}{1968}), \bibinfo{month}{Aug.}
    \bibinfo{title}{Two-dimensional stellar collisions}.
    \bibinfo{journal}{{\em Astrophysical Journal}} \bibinfo{volume}{153}:
      \bibinfo{pages}{633}.
    
    \bibtype{Article}%
    \bibitem[Deheuvels et al.(2022)]{deheuvels2022a}
    \bibinfo{author}{Deheuvels S}, \bibinfo{author}{Ballot J},
      \bibinfo{author}{Gehan C} and  \bibinfo{author}{Mosser B}
      (\bibinfo{year}{2022}), \bibinfo{month}{Mar.}
    \bibinfo{title}{Seismic signature of electron degeneracy in the core of red
      giants: {Hints} for mass transfer between close red-giant companions}.
    \bibinfo{journal}{{\em Astronomy and Astrophysics}} \bibinfo{volume}{659}:
      \bibinfo{pages}{A106}.
    
    \bibtype{Article}%
    \bibitem[Di~Carlo et al.(2019)]{dicarlo2019a}
    \bibinfo{author}{Di~Carlo UN}, \bibinfo{author}{Giacobbo N},
      \bibinfo{author}{Mapelli M}, \bibinfo{author}{Pasquato M},
      \bibinfo{author}{Spera M}, \bibinfo{author}{Wang L} and
      \bibinfo{author}{Haardt F} (\bibinfo{year}{2019}), \bibinfo{month}{Aug.}
    \bibinfo{title}{Merging black holes in young star clusters}.
    \bibinfo{journal}{{\em Monthly Notices of the Royal Astronomical Society}}
      \bibinfo{volume}{487} (\bibinfo{number}{2}): \bibinfo{pages}{2947--2960}.
    
    \bibtype{Article}%
    \bibitem[Dray and Tout(2007)]{dray2007a}
    \bibinfo{author}{Dray LM} and  \bibinfo{author}{Tout CA}
      (\bibinfo{year}{2007}), \bibinfo{month}{Mar.}
    \bibinfo{title}{On rejuvenation in massive binary systems}.
    \bibinfo{journal}{{\em Monthly Notices of the Royal Astronomical Society}}
      \bibinfo{volume}{376}: \bibinfo{pages}{61--70}.
    
    \bibtype{Article}%
    \bibitem[Dvořáková et al.(2024)]{dvorakova2024a}
    \bibinfo{author}{Dvořáková N}, \bibinfo{author}{Korčáková D},
      \bibinfo{author}{Dinnbier F} and  \bibinfo{author}{Kroupa P}
      (\bibinfo{year}{2024}), \bibinfo{month}{Sep.}
    \bibinfo{title}{The mass distribution of stellar mergers: {A} new scenario for
      several {FS} {CMa} stars}.
    \bibinfo{journal}{{\em Astronomy and Astrophysics}} \bibinfo{volume}{689}:
      \bibinfo{pages}{A234}.
    
    \bibtype{Article}%
    \bibitem[Ferrario et al.(2009)]{ferrario2009a}
    \bibinfo{author}{Ferrario L}, \bibinfo{author}{Pringle JE},
      \bibinfo{author}{Tout CA} and  \bibinfo{author}{Wickramasinghe DT}
      (\bibinfo{year}{2009}), \bibinfo{month}{Nov.}
    \bibinfo{title}{The origin of magnetism on the upper main sequence}.
    \bibinfo{journal}{{\em Monthly Notices of the Royal Astronomical Society}}
      \bibinfo{volume}{400}: \bibinfo{pages}{L71--L74}.
    
    \bibtype{Article}%
    \bibitem[Ferraro et al.(2023)]{ferraro2023a}
    \bibinfo{author}{Ferraro FR}, \bibinfo{author}{Mucciarelli A},
      \bibinfo{author}{Lanzoni B}, \bibinfo{author}{Pallanca C},
      \bibinfo{author}{Cadelano M}, \bibinfo{author}{Billi A},
      \bibinfo{author}{Sills A}, \bibinfo{author}{Vesperini E},
      \bibinfo{author}{Dalessandro E}, \bibinfo{author}{Beccari G},
      \bibinfo{author}{Monaco L} and  \bibinfo{author}{Mateo M}
      (\bibinfo{year}{2023}), \bibinfo{month}{May}.
    \bibinfo{title}{Fast rotating blue stragglers prefer loose clusters}.
    \bibinfo{journal}{{\em Nature Communications}} \bibinfo{volume}{14}:
      \bibinfo{pages}{2584}.
    
    \bibtype{Article}%
    \bibitem[Freitag and Benz(2005)]{freitag2005a}
    \bibinfo{author}{Freitag M} and  \bibinfo{author}{Benz W}
      (\bibinfo{year}{2005}), \bibinfo{month}{Apr.}
    \bibinfo{title}{A comprehensive set of simulations of high-velocity collisions
      between main-sequence stars}.
    \bibinfo{journal}{{\em Monthly Notices of the Royal Astronomical Society}}
      \bibinfo{volume}{358}: \bibinfo{pages}{1133--1158}.
    
    \bibtype{Article}%
    \bibitem[Frost et al.(2024)]{frost2024a}
    \bibinfo{author}{Frost AJ}, \bibinfo{author}{Sana H}, \bibinfo{author}{Mahy L},
      \bibinfo{author}{Wade G}, \bibinfo{author}{Barron J},
      \bibinfo{author}{Bouquin JBL}, \bibinfo{author}{Mérand A},
      \bibinfo{author}{Schneider FRN}, \bibinfo{author}{Shenar T},
      \bibinfo{author}{Barbá RH}, \bibinfo{author}{Bowman DM},
      \bibinfo{author}{Fabry M}, \bibinfo{author}{Farhang A},
      \bibinfo{author}{Marchant P}, \bibinfo{author}{Morrell NI} and
      \bibinfo{author}{Smoker JV} (\bibinfo{year}{2024}).
    \bibinfo{title}{A magnetic massive star has experienced a stellar merger}.
    \bibinfo{journal}{{\em Science}} \bibinfo{volume}{384}
      (\bibinfo{number}{6692}): \bibinfo{pages}{214--217}.
    
    \bibtype{Article}%
    \bibitem[Gaburov et al.(2008)]{gaburov2008a}
    \bibinfo{author}{Gaburov E}, \bibinfo{author}{Lombardi JC} and
      \bibinfo{author}{Portegies~Zwart S} (\bibinfo{year}{2008}),
      \bibinfo{month}{Jan.}
    \bibinfo{title}{Mixing in massive stellar mergers}.
    \bibinfo{journal}{{\em Monthly Notices of the Royal Astronomical Society}}
      \bibinfo{volume}{383}: \bibinfo{pages}{L5--L9}.
    
    \bibtype{Article}%
    \bibitem[Glanz and Perets(2018)]{glanz2018a}
    \bibinfo{author}{Glanz H} and  \bibinfo{author}{Perets HB}
      (\bibinfo{year}{2018}), \bibinfo{month}{Jul.}
    \bibinfo{title}{Efficient common-envelope ejection through dust-driven winds}.
    \bibinfo{journal}{{\em Monthly Notices of the Royal Astronomical Society}}
      \bibinfo{volume}{478} (\bibinfo{number}{1}): \bibinfo{pages}{L12--L17}.
    
    \bibtype{Article}%
    \bibitem[Glanz and Perets(2021)]{glanz2021b}
    \bibinfo{author}{Glanz H} and  \bibinfo{author}{Perets HB}
      (\bibinfo{year}{2021}), \bibinfo{month}{Jan.}
    \bibinfo{title}{Simulations of common envelope evolution in triple systems:
      circumstellar case}.
    \bibinfo{journal}{{\em Monthly Notices of the Royal Astronomical Society}}
      \bibinfo{volume}{500} (\bibinfo{number}{2}): \bibinfo{pages}{1921--1932}.
    
    \bibtype{Article}%
    \bibitem[Glebbeek and Pols(2008)]{glebbeek2008a}
    \bibinfo{author}{Glebbeek E} and  \bibinfo{author}{Pols OR}
      (\bibinfo{year}{2008}), \bibinfo{month}{Sep.}
    \bibinfo{title}{Evolution of stellar collision products in open clusters. {II}.
      {A} grid of low-mass collisions}.
    \bibinfo{journal}{{\em Astronomy and Astrophysics}} \bibinfo{volume}{488}:
      \bibinfo{pages}{1017--1025}.
    
    \bibtype{Article}%
    \bibitem[Glebbeek et al.(2013)]{glebbeek2013a}
    \bibinfo{author}{Glebbeek E}, \bibinfo{author}{Gaburov E},
      \bibinfo{author}{Portegies~Zwart S} and  \bibinfo{author}{Pols OR}
      (\bibinfo{year}{2013}), \bibinfo{month}{Oct.}
    \bibinfo{title}{Structure and evolution of high-mass stellar mergers}.
    \bibinfo{journal}{{\em Monthly Notices of the Royal Astronomical Society}}
      \bibinfo{volume}{434}: \bibinfo{pages}{3497--3510}.
    
    \bibtype{Article}%
    \bibitem[Harris and Tricco(2023)]{harris2023a}
    \bibinfo{author}{Harris A} and  \bibinfo{author}{Tricco T}
      (\bibinfo{year}{2023}), \bibinfo{month}{Jun.}
    \bibinfo{title}{Sarracen: a {Python} package for analysis and visualization of
      smoothed particle hydrodynamics data}.
    \bibinfo{journal}{{\em The Journal of Open Source Software}}
      \bibinfo{volume}{8}: \bibinfo{pages}{5263}.
    
    \bibtype{Article}%
    \bibitem[Hartmann et al.(2016)]{hartmann2016a}
    \bibinfo{author}{Hartmann L}, \bibinfo{author}{Herczeg G} and
      \bibinfo{author}{Calvet N} (\bibinfo{year}{2016}), \bibinfo{month}{Sep.}
    \bibinfo{title}{Accretion onto pre-main-sequence stars}.
    \bibinfo{journal}{{\em Annual Review of Astron and Astrophys}}
      \bibinfo{volume}{54}: \bibinfo{pages}{135--180}.
    
    \bibtype{Article}%
    \bibitem[Heggie(1975)]{heggie1975a}
    \bibinfo{author}{Heggie DC} (\bibinfo{year}{1975}), \bibinfo{month}{Dec.}
    \bibinfo{title}{Binary evolution in stellar dynamics.}
    \bibinfo{journal}{{\em Monthly Notices of the Royal Astronomical Society}}
      \bibinfo{volume}{173}: \bibinfo{pages}{729--787}.
    
    \bibtype{Article}%
    \bibitem[Hellings(1983)]{hellings1983a}
    \bibinfo{author}{Hellings P} (\bibinfo{year}{1983}), \bibinfo{month}{Oct.}
    \bibinfo{title}{Phenomenological study of massive accretion stars}.
    \bibinfo{journal}{{\em Astrophysics and Space Science}} \bibinfo{volume}{96}
      (\bibinfo{number}{1}): \bibinfo{pages}{37--54}.
    
    \bibtype{Article}%
    \bibitem[Henneco et al.(2024{\natexlab{a}})]{henneco2024b}
    \bibinfo{author}{Henneco J}, \bibinfo{author}{Schneider FRN},
      \bibinfo{author}{Hekker S} and  \bibinfo{author}{Aerts C}
      (\bibinfo{year}{2024}{\natexlab{a}}), \bibinfo{month}{Oct.}
    \bibinfo{title}{Merger seismology: distinguishing massive merger products from
      genuine single stars using asteroseismology}.
    \bibinfo{journal}{{\em Astronomy and Astrophysics}} \bibinfo{volume}{690}
      \bibinfo{pages}{A65}.
    
    \bibtype{Article}%
    \bibitem[Henneco et al.(2024{\natexlab{b}})]{henneco2024a}
    \bibinfo{author}{Henneco J}, \bibinfo{author}{Schneider FRN} and
      \bibinfo{author}{Laplace E} (\bibinfo{year}{2024}{\natexlab{b}}),
      \bibinfo{month}{Feb.}
    \bibinfo{title}{Contact tracing of binary stars: {Pathways} to stellar
      mergers}.
    \bibinfo{journal}{{\em Astronomy and Astrophysics}} \bibinfo{volume}{682}:
      \bibinfo{pages}{A169}.
    
    \bibtype{Article}%
    \bibitem[Hills and Day(1976)]{hills1976a}
    \bibinfo{author}{Hills JG} and  \bibinfo{author}{Day CA}
      (\bibinfo{year}{1976}), \bibinfo{month}{Feb.}
    \bibinfo{title}{Stellar collisions in globular clusters}.
    \bibinfo{journal}{{\em Astrophysics Letters}} \bibinfo{volume}{17}:
    \bibinfo{pages}{87}.
    
    \bibtype{Article}%
    \bibitem[Hirai et al.(2021)]{hirai2021a}
    \bibinfo{author}{Hirai R}, \bibinfo{author}{Podsiadlowski P},
      \bibinfo{author}{Owocki SP}, \bibinfo{author}{Schneider FRN} and
      \bibinfo{author}{Smith N} (\bibinfo{year}{2021}), \bibinfo{month}{May}.
    \bibinfo{title}{Simulating the formation of η {Carinae}'s surrounding nebula
      through unstable triple evolution and stellar merger-induced eruption}.
    \bibinfo{journal}{{\em Monthly Notices of the Royal Astronomical Society}}
      \bibinfo{volume}{503} (\bibinfo{number}{3}): \bibinfo{pages}{4276--4296}.
    
    \bibtype{Article}%
    \bibitem[Howitt et al.(2020)]{howitt2020a}
    \bibinfo{author}{Howitt G}, \bibinfo{author}{Stevenson S},
      \bibinfo{author}{Vigna-Gómez Ar}, \bibinfo{author}{Justham S},
      \bibinfo{author}{Ivanova N}, \bibinfo{author}{Woods TE},
      \bibinfo{author}{Neijssel CJ} and  \bibinfo{author}{Mandel I}
      (\bibinfo{year}{2020}), \bibinfo{month}{Mar.}
    \bibinfo{title}{Luminous {Red} {Novae}: population models and future
      prospects}.
    \bibinfo{journal}{{\em Monthly Notices of the Royal Astronomical Society}}
      \bibinfo{volume}{492} (\bibinfo{number}{3}): \bibinfo{pages}{3229--3240}.
    
    \bibtype{Article}%
    \bibitem[Humphreys and Davidson(1979)]{humphreys1979a}
    \bibinfo{author}{Humphreys RM} and  \bibinfo{author}{Davidson K}
      (\bibinfo{year}{1979}), \bibinfo{month}{Sep.}
    \bibinfo{title}{Studies of luminous stars in nearby galaxies. {III} -
      {Comments} on the evolution of the most massive stars in the {Milky} {Way}
      and the {Large} {Magellanic} {Cloud}}.
    \bibinfo{journal}{{\em Astrophysical Journal}} \bibinfo{volume}{232}:
      \bibinfo{pages}{409--420}.
    
    \bibtype{Article}%
    \bibitem[Hut(1980)]{hut1980a}
    \bibinfo{author}{Hut P} (\bibinfo{year}{1980}), \bibinfo{month}{Dec.}
    \bibinfo{title}{Stability of tidal equilibrium}.
    \bibinfo{journal}{{\em Astronomy and Astrophysics}} \bibinfo{volume}{92}:
      \bibinfo{pages}{167--170}.
    
    \bibtype{Article}%
    \bibitem[Iaconi and De~Marco(2019)]{iaconi2019a}
    \bibinfo{author}{Iaconi R} and  \bibinfo{author}{De~Marco O}
      (\bibinfo{year}{2019}), \bibinfo{month}{Dec.}
    \bibinfo{title}{Speaking with one voice: simulations and observations discuss
      the common envelope α parameter}.
    \bibinfo{journal}{{\em Monthly Notices of the Royal Astronomical Society}}
      \bibinfo{volume}{490} (\bibinfo{number}{2}): \bibinfo{pages}{2550--2566}.
    
    \bibtype{Article}%
    \bibitem[Ivanova et al.(2013{\natexlab{a}})]{ivanova2013a}
    \bibinfo{author}{Ivanova N}, \bibinfo{author}{Justham S}, \bibinfo{author}{Chen
      X}, \bibinfo{author}{De~Marco O}, \bibinfo{author}{Fryer CL},
      \bibinfo{author}{Gaburov E}, \bibinfo{author}{Ge H},
      \bibinfo{author}{Glebbeek E}, \bibinfo{author}{Han Z}, \bibinfo{author}{Li
      XD}, \bibinfo{author}{Lu G}, \bibinfo{author}{Marsh T},
      \bibinfo{author}{Podsiadlowski P}, \bibinfo{author}{Potter A},
      \bibinfo{author}{Soker N}, \bibinfo{author}{Taam R}, \bibinfo{author}{Tauris
      TM}, \bibinfo{author}{van~den Heuvel EPJ} and  \bibinfo{author}{Webbink RF}
      (\bibinfo{year}{2013}{\natexlab{a}}), \bibinfo{month}{Feb.}
    \bibinfo{title}{Common envelope evolution: where we stand and how we can move
      forward}.
    \bibinfo{journal}{{\em Astronomy and Astrophysics Reviews}}
      \bibinfo{volume}{21}: \bibinfo{pages}{59}.
    
    \bibtype{Article}%
    \bibitem[Ivanova et al.(2013{\natexlab{b}})]{ivanova2013b}
    \bibinfo{author}{Ivanova N}, \bibinfo{author}{Justham S},
      \bibinfo{author}{Nandez JLA} and  \bibinfo{author}{Lombardi JC}
      (\bibinfo{year}{2013}{\natexlab{b}}), \bibinfo{month}{Jan.}
    \bibinfo{title}{Identification of the long-sought common-envelope events}.
    \bibinfo{journal}{{\em Science}} \bibinfo{volume}{339}: \bibinfo{pages}{433}.
    
    \bibtype{Article}%
    \bibitem[Jacoby et al.(2021)]{jacoby2021a}
    \bibinfo{author}{Jacoby GH}, \bibinfo{author}{Hillwig TC},
      \bibinfo{author}{Jones D}, \bibinfo{author}{Martin K},
      \bibinfo{author}{De~Marco O}, \bibinfo{author}{Kronberger M},
      \bibinfo{author}{Hurowitz JL}, \bibinfo{author}{Crocker AF} and
      \bibinfo{author}{Dey J} (\bibinfo{year}{2021}), \bibinfo{month}{Oct.}
    \bibinfo{title}{Binary central stars of planetary nebulae identified with
      {Kepler}/{K2}}.
    \bibinfo{journal}{{\em Monthly Notices of the Royal Astronomical Society}}
      \bibinfo{volume}{506} (\bibinfo{number}{4}): \bibinfo{pages}{5223--5246}.
    
    \bibtype{Article}%
    \bibitem[Joss et al.(1987)]{joss1987a}
    \bibinfo{author}{Joss PC}, \bibinfo{author}{Rappaport S} and
      \bibinfo{author}{Lewis W} (\bibinfo{year}{1987}), \bibinfo{month}{Aug.}
    \bibinfo{title}{The {Core} {Mass}--{Radius} {Relation} for {Giants}: {A} {New}
      {Test} of {Stellar} {Evolution} {Theory}}.
    \bibinfo{journal}{{\em The Astrophysical Journal}} \bibinfo{volume}{319}:
      \bibinfo{pages}{180}.
    
    \bibtype{Article}%
    \bibitem[Justham et al.(2014)]{justham2014a}
    \bibinfo{author}{Justham S}, \bibinfo{author}{Podsiadlowski P} and
      \bibinfo{author}{Vink JS} (\bibinfo{year}{2014}), \bibinfo{month}{Dec.}
    \bibinfo{title}{Luminous blue variables and superluminous supernovae from
      binary mergers}.
    \bibinfo{journal}{{\em Astrophysical Journal}} \bibinfo{volume}{796}:
      \bibinfo{pages}{121}.
    
    \bibtype{Article}%
    \bibitem[Kim and Kim(2007)]{kim2007a}
    \bibinfo{author}{Kim H} and  \bibinfo{author}{Kim WT} (\bibinfo{year}{2007}),
      \bibinfo{month}{Aug.}
    \bibinfo{title}{Dynamical friction of a circular-orbit perturber in a gaseous
      medium}.
    \bibinfo{journal}{{\em Astrophysical Journal}} \bibinfo{volume}{665}
      (\bibinfo{number}{1}): \bibinfo{pages}{432--444}.
    
    \bibtype{Article}%
    \bibitem[Kippenhahn and Meyer-Hofmeister(1977)]{kippenhahn1977a}
    \bibinfo{author}{Kippenhahn R} and  \bibinfo{author}{Meyer-Hofmeister E}
      (\bibinfo{year}{1977}), \bibinfo{month}{Jan.}
    \bibinfo{title}{On the radii of accreting main sequence stars}.
    \bibinfo{journal}{{\em Astronomy and Astrophysics}} \bibinfo{volume}{54}:
      \bibinfo{pages}{539--542}.
    
    \bibtype{Book}%
    \bibitem[Kippenhahn et al.(2012)]{kippenhahn2012a}
    \bibinfo{author}{Kippenhahn R}, \bibinfo{author}{Weigert A} and
      \bibinfo{author}{Weiss A} (\bibinfo{year}{2012}).
    \bibinfo{title}{Stellar structure and evolution},
      \bibinfo{publisher}{Springer-Verlag}, \bibinfo{address}{Berlin Heidelberg}.
    \bibinfo{comment}{ISBN} \bibinfo{isbn}{978-3-642-30255-8}.

    \bibtype{Article}%
    \bibitem[Kiuchi et al.(2024)]{kiuchi2024a}
    \bibinfo{author}{Kiuchi K}, \bibinfo{author}{Reboul-Salze A},
      \bibinfo{author}{Shibata M} and  \bibinfo{author}{Sekiguchi Y}
      (\bibinfo{year}{2024}), \bibinfo{month}{Mar.}
    \bibinfo{title}{A large-scale magnetic field produced by a solar-like dynamo in
      binary neutron star mergers}.
    \bibinfo{journal}{{\em Nature Astronomy}} \bibinfo{volume}{8}:
      \bibinfo{pages}{298--307}.
    
    \bibtype{Article}%
    \bibitem[Kozai(1962)]{kozai1962a}
    \bibinfo{author}{Kozai Y} (\bibinfo{year}{1962}), \bibinfo{month}{Nov.}
    \bibinfo{title}{Secular perturbations of asteroids with high inclination and
      eccentricity}.
    \bibinfo{journal}{{\em The Astronomical Journal}} \bibinfo{volume}{67}:
      \bibinfo{pages}{591--598}.
    
    \bibtype{Article}%
    \bibitem[Kramer et al.(2020)]{kramer2020a}
    \bibinfo{author}{Kramer M}, \bibinfo{author}{Schneider FRN},
      \bibinfo{author}{Ohlmann ST}, \bibinfo{author}{Geier S},
      \bibinfo{author}{Schaffenroth V}, \bibinfo{author}{Pakmor R} and
      \bibinfo{author}{Röpke FK} (\bibinfo{year}{2020}), \bibinfo{month}{Oct.}
    \bibinfo{title}{Formation of {sdB}-stars via common envelope ejection by
      substellar companions}.
    \bibinfo{journal}{{\em Astronomy and Astrophysics}} \bibinfo{volume}{642}:
      \bibinfo{pages}{A97}.
    
    \bibtype{Book}%
    \bibitem[Landau and Lifshitz(1959)]{landaulifshitz6eng}
    \bibinfo{author}{Landau LD} and  \bibinfo{author}{Lifshitz EM}
      (\bibinfo{year}{1959}).
    \bibinfo{title}{Fluid mechanics}, \bibinfo{series}{Course of theoretical
      physics}, \bibinfo{volume}{6}, \bibinfo{publisher}{Pergamon Press},
    
    \bibtype{Article}%
    \bibitem[Langer(2012)]{langer2012a}
    \bibinfo{author}{Langer N} (\bibinfo{year}{2012}), \bibinfo{month}{Sep.}
    \bibinfo{title}{Presupernova evolution of massive single and binary stars}.
    \bibinfo{journal}{{\em Annual Review of Astronomy and Astrophysics}}
      \bibinfo{volume}{50}: \bibinfo{pages}{107--164}.
    
    \bibtype{Misc}%
    \bibitem[Lau et al.(2022{\natexlab{a}})]{lau2022c}
    \bibinfo{author}{Lau MYM}, \bibinfo{author}{Cantiello M},
      \bibinfo{author}{Jermyn AS}, \bibinfo{author}{MacLeod M},
      \bibinfo{author}{Mandel I} and  \bibinfo{author}{Price DJ}
      (\bibinfo{year}{2022}{\natexlab{a}}), \bibinfo{month}{Oct.}
    \bibinfo{title}{Hot {Jupiter} engulfment by a red giant in {3D} hydrodynamics}.
    \bibinfo{url}{\url{https://doi.org/10.48550/arXiv.2210.15848}}.
    
    \bibtype{Article}%
    \bibitem[Lau et al.(2022{\natexlab{b}})]{lau2022a}
    \bibinfo{author}{Lau MYM}, \bibinfo{author}{Hirai R},
      \bibinfo{author}{González-Bolívar M}, \bibinfo{author}{Price DJ},
      \bibinfo{author}{De~Marco O} and  \bibinfo{author}{Mandel I}
      (\bibinfo{year}{2022}{\natexlab{b}}), \bibinfo{month}{Jun.}
    \bibinfo{title}{Common envelopes in massive stars: {Towards} the role of
      radiation pressure and recombination energy in ejecting red supergiant
      envelopes}.
    \bibinfo{journal}{{\em Monthly Notices of the Royal Astronomical Society}}
      \bibinfo{volume}{512}:
      \bibinfo{pages}{5462--5480}.
    
    \bibtype{Article}%
    \bibitem[Lau et al.(2024)]{lau2024a}
    \bibinfo{author}{Lau MYM}, \bibinfo{author}{Hirai R}, \bibinfo{author}{Mandel
      I} and  \bibinfo{author}{Tout CA} (\bibinfo{year}{2024}),
      \bibinfo{month}{May}.
    \bibinfo{title}{Expansion of {Accreting} {Main}-sequence {Stars} during {Rapid}
      {Mass} {Transfer}}.
    \bibinfo{journal}{{\em The Astrophysical Journal}} \bibinfo{volume}{966}:
      \bibinfo{pages}{L7}.
    
    \bibtype{Article}%
    \bibitem[Leonard and Livio(1995)]{leonard1995a}
    \bibinfo{author}{Leonard PJT} and  \bibinfo{author}{Livio M}
      (\bibinfo{year}{1995}), \bibinfo{month}{Jul.}
    \bibinfo{title}{The rotational rates of blue stragglers produced by physical
      stellar collisions}.
    \bibinfo{journal}{{\em Astrophysical Journal, Letters}} \bibinfo{volume}{447}:
      \bibinfo{pages}{L121}.
    
    \bibtype{Article}%
    \bibitem[Li et al.(2022)]{li2022a}
    \bibinfo{author}{Li Y}, \bibinfo{author}{Bedding TR}, \bibinfo{author}{Murphy
      SJ}, \bibinfo{author}{Stello D}, \bibinfo{author}{Chen Y},
      \bibinfo{author}{Huber D}, \bibinfo{author}{Joyce M}, \bibinfo{author}{Marks
      D}, \bibinfo{author}{Zhang X}, \bibinfo{author}{Bi S},
      \bibinfo{author}{Colman IL}, \bibinfo{author}{Hayden MR},
      \bibinfo{author}{Hey DR}, \bibinfo{author}{Li G}, \bibinfo{author}{Montet
      BT}, \bibinfo{author}{Sharma S} and  \bibinfo{author}{Wu Y}
      (\bibinfo{year}{2022}), \bibinfo{month}{Apr.}
    \bibinfo{title}{Discovery of post-mass-transfer helium-burning red giants using
      asteroseismology}.
    \bibinfo{journal}{{\em Nature Astronomy}} \bibinfo{volume}{6}:
      \bibinfo{pages}{673--680}.
    
    \bibtype{Article}%
    \bibitem[Lidov(1962)]{lidov1962a}
    \bibinfo{author}{Lidov ML} (\bibinfo{year}{1962}), \bibinfo{month}{Oct.}
    \bibinfo{title}{The evolution of orbits of artificial satellites of planets
      under the action of gravitational perturbations of external bodies}.
    \bibinfo{journal}{{\em Planetary and Space Science}} \bibinfo{volume}{9}:
      \bibinfo{pages}{719--759}.
    
    \bibtype{Article}%
    \bibitem[Lombardi et al.(1996)]{lombardi1996a}
    \bibinfo{author}{Lombardi Jr. JC}, \bibinfo{author}{Rasio FA} and
      \bibinfo{author}{Shapiro SL} (\bibinfo{year}{1996}), \bibinfo{month}{Sep.}
    \bibinfo{title}{Collisions of main-sequence stars and the formation of blue
      stragglers in globular clusters}.
    \bibinfo{journal}{{\em Astrophysical Journal}} \bibinfo{volume}{468}:
      \bibinfo{pages}{797}.
    
    \bibtype{Article}%
    \bibitem[Lombardi et al.(2002)]{lombardi2002a}
    \bibinfo{author}{Lombardi Jr. JC}, \bibinfo{author}{Warren JS},
      \bibinfo{author}{Rasio FA}, \bibinfo{author}{Sills A} and
      \bibinfo{author}{Warren AR} (\bibinfo{year}{2002}), \bibinfo{month}{Apr.}
    \bibinfo{title}{Stellar collisions and the interior structure of blue
      stragglers}.
    \bibinfo{journal}{{\em Astrophysical Journal}} \bibinfo{volume}{568}:
      \bibinfo{pages}{939--953}.
    
    \bibtype{Article}%
    \bibitem[Marchant and Bodensteiner(2024)]{marchant2024a}
    \bibinfo{author}{Marchant P} and  \bibinfo{author}{Bodensteiner J}
      (\bibinfo{year}{2024}), \bibinfo{month}{Sep.}
    \bibinfo{title}{The {Evolution} of {Massive} {Binary} {Stars}}.
    \bibinfo{journal}{{\em Annual Review of Astronomy and Astrophysics}}
      \bibinfo{volume}{62}: \bibinfo{pages}{21--61}.
    
    \bibtype{Article}%
    \bibitem[Mason et al.(2010)]{mason2010b}
    \bibinfo{author}{Mason E}, \bibinfo{author}{Diaz M}, \bibinfo{author}{Williams
      RE}, \bibinfo{author}{Preston G} and  \bibinfo{author}{Bensby T}
      (\bibinfo{year}{2010}), \bibinfo{month}{Jun.}
    \bibinfo{title}{The peculiar nova {V1309} {Scorpii}/nova {Scorpii} 2008. {A}
      candidate twin of {V838} {Monocerotis}}.
    \bibinfo{journal}{{\em Astronomy and Astrophysics}} \bibinfo{volume}{516}:
      \bibinfo{pages}{A108}.
    
    \bibtype{Article}%
    \bibitem[Mathis(1967)]{mathis1967a}
    \bibinfo{author}{Mathis JS} (\bibinfo{year}{1967}), \bibinfo{month}{Mar.}
    \bibinfo{title}{Nuclear reactions during stellar collisions}.
    \bibinfo{journal}{{\em Astrophysical Journal}} \bibinfo{volume}{147}:
      \bibinfo{pages}{1050}.
    
    \bibtype{Article}%
    \bibitem[Matsumoto and Metzger(2022)]{matsumoto2022a}
    \bibinfo{author}{Matsumoto T} and  \bibinfo{author}{Metzger BD}
      (\bibinfo{year}{2022}), \bibinfo{month}{Oct.}
    \bibinfo{title}{Light-curve model for luminous red novae and inferences about
      the ejecta of stellar mergers}.
    \bibinfo{journal}{{\em Astrophysical Journal}} \bibinfo{volume}{938}
      (\bibinfo{number}{1}): \bibinfo{pages}{5}.
    
    \bibtype{Article}%
    \bibitem[Menon and Heger(2017)]{menon2017a}
    \bibinfo{author}{Menon A} and  \bibinfo{author}{Heger A}
      (\bibinfo{year}{2017}), \bibinfo{month}{Aug.}
    \bibinfo{title}{The quest for blue supergiants: binary merger models for the
      evolution of the progenitor of {SN} {1987A}.}
    \bibinfo{journal}{{\em Monthly Notices of the Royal Astronomical Society}}
      \bibinfo{volume}{469}:
      \bibinfo{pages}{4649--4664}.
    
    \bibtype{Article}%
    \bibitem[Menon et al.(2024)]{menon2024a}
    \bibinfo{author}{Menon A}, \bibinfo{author}{Ercolino A},
      \bibinfo{author}{Urbaneja MA}, \bibinfo{author}{Lennon DJ},
      \bibinfo{author}{Herrero A}, \bibinfo{author}{Hirai R},
      \bibinfo{author}{Langer N}, \bibinfo{author}{Schootemeijer A},
      \bibinfo{author}{Chatzopoulos E}, \bibinfo{author}{Frank J} and
      \bibinfo{author}{Shiber S} (\bibinfo{year}{2024}), \bibinfo{month}{Mar.}
    \bibinfo{title}{Evidence for {Evolved} {Stellar} {Binary} {Mergers} in
      {Observed} {B}-type {Blue} {Supergiants}}.
    \bibinfo{journal}{{\em The Astrophysical Journal}} \bibinfo{volume}{963}
      (\bibinfo{number}{2}): \bibinfo{pages}{L42}.
    
    \bibtype{Article}%
    \bibitem[Meyer and Meyer-Hofmeister(1979)]{meyer1979a}
    \bibinfo{author}{Meyer F} and  \bibinfo{author}{Meyer-Hofmeister E}
      (\bibinfo{year}{1979}), \bibinfo{month}{Sep.}
    \bibinfo{title}{Formation of cataclysmic binaries through common envelope
      evolution}.
    \bibinfo{journal}{{\em Astronomy and Astrophysics}} \bibinfo{volume}{78}:
      \bibinfo{pages}{167--176}.
    
    \bibtype{Article}%
    \bibitem[Miszalski et al.(2009)]{miszalski2009a}
    \bibinfo{author}{Miszalski B}, \bibinfo{author}{Acker A},
      \bibinfo{author}{Moffat AFJ}, \bibinfo{author}{Parker QA} and
      \bibinfo{author}{Udalski A} (\bibinfo{year}{2009}), \bibinfo{month}{Mar.}
    \bibinfo{title}{Binary planetary nebulae nuclei towards the {Galactic} bulge.
      {I}. {Sample} discovery, period distribution, and binary fraction}.
    \bibinfo{journal}{{\em Astronomy and Astrophysics}} \bibinfo{volume}{496}:
      \bibinfo{pages}{813--825}.
    
    \bibtype{Article}%
    \bibitem[Moe and Di~Stefano(2017)]{moe2017a}
    \bibinfo{author}{Moe M} and  \bibinfo{author}{Di~Stefano R}
      (\bibinfo{year}{2017}), \bibinfo{month}{Jun.}
    \bibinfo{title}{Mind your {Ps} and {Qs}: {The} interrelation between period
      ({P}) and mass-ratio ({Q}) distributions of binary stars}.
    \bibinfo{journal}{{\em Astrophysical Journal, Supplement}}
      \bibinfo{volume}{230} (\bibinfo{number}{2}): \bibinfo{pages}{15}.
    
    \bibtype{Article}%
    \bibitem[Moreno et al.(2022)]{moreno2022a}
    \bibinfo{author}{Moreno MM}, \bibinfo{author}{Schneider FRN},
      \bibinfo{author}{Röpke FK}, \bibinfo{author}{Ohlmann ST},
      \bibinfo{author}{Pakmor R}, \bibinfo{author}{Podsiadlowski P} and
      \bibinfo{author}{Sand C} (\bibinfo{year}{2022}), \bibinfo{month}{Nov.}
    \bibinfo{title}{From {3D} hydrodynamic simulations of common-envelope
      interaction to gravitational-wave mergers}.
    \bibinfo{journal}{{\em Astronomy and Astrophysics}} \bibinfo{volume}{667}:
      \bibinfo{pages}{A72}.
    
    \bibtype{Article}%
    \bibitem[Munari et al.(2002)]{munari2002a}
    \bibinfo{author}{Munari U}, \bibinfo{author}{Henden A}, \bibinfo{author}{Kiyota
      S}, \bibinfo{author}{Laney D}, \bibinfo{author}{Marang F},
      \bibinfo{author}{Zwitter T}, \bibinfo{author}{Corradi RLM},
      \bibinfo{author}{Desidera S}, \bibinfo{author}{Marrese PM},
      \bibinfo{author}{Giro E}, \bibinfo{author}{Boschi F} and
      \bibinfo{author}{Schwartz MB} (\bibinfo{year}{2002}), \bibinfo{month}{Jul.}
    \bibinfo{title}{The mysterious eruption of {V838} {Mon}}.
    \bibinfo{journal}{{\em Astronomy and Astrophysics}} \bibinfo{volume}{389}:
      \bibinfo{pages}{L51--L56}.
    
    \bibtype{Article}%
    \bibitem[Nakano et al.(2008)]{nakano2008a}
    \bibinfo{author}{Nakano S}, \bibinfo{author}{Nishiyama K},
      \bibinfo{author}{Kabashima F}, \bibinfo{author}{Sakurai Y},
      \bibinfo{author}{Jacques C}, \bibinfo{author}{Pimentel E},
      \bibinfo{author}{Chekhovich D}, \bibinfo{author}{Korotkiy S},
      \bibinfo{author}{Kryachko T} and  \bibinfo{author}{Samus NN}
      (\bibinfo{year}{2008}), \bibinfo{month}{Sep.}
    \bibinfo{title}{V1309 {Scorpii} = {Nova} {Scorpii} 2008}.
    \bibinfo{journal}{{\em International Astronomical Union Circular}}
      \bibinfo{volume}{8972}: \bibinfo{pages}{1}.
    
    
    \bibtype{Article}%
    \bibitem[Nandez et al.(2014)]{nandez2014a}
    \bibinfo{author}{Nandez JLA}, \bibinfo{author}{Ivanova N} and
      \bibinfo{author}{Lombardi Jr. JC} (\bibinfo{year}{2014}),
      \bibinfo{month}{May}.
    \bibinfo{title}{V1309 {Sco} – understanding a merger}.
    \bibinfo{journal}{{\em Astrophysical Journal}} \bibinfo{volume}{786}:
      \bibinfo{pages}{39}.
    
    \bibtype{Inproceedings}%
    \bibitem[Offner et al.(2023)]{offner2023a}
    \bibinfo{author}{Offner SSR}, \bibinfo{author}{Moe M}, \bibinfo{author}{Kratter
      KM}, \bibinfo{author}{Sadavoy SI}, \bibinfo{author}{Jensen ELN} and
      \bibinfo{author}{Tobin JJ} (\bibinfo{year}{2023}), \bibinfo{month}{Jul.},
      \bibinfo{title}{The origin and evolution of multiple star systems},
      \bibinfo{editor}{Inutsuka S}, \bibinfo{editor}{Aikawa Y},
      \bibinfo{editor}{Muto T}, \bibinfo{editor}{Tomida K} and
      \bibinfo{editor}{Tamura M}, (Eds.), \bibinfo{booktitle}{Protostars and
      planets {VII}}, \bibinfo{series}{Astronomical society of the pacific
      conference series}, \bibinfo{volume}{534}, pp. \bibinfo{pages}{275}.
    
    \bibtype{Article}%
    \bibitem[Ohlmann et al.(2016{\natexlab{a}})]{ohlmann2016a}
    \bibinfo{author}{Ohlmann ST}, \bibinfo{author}{Röpke FK},
      \bibinfo{author}{Pakmor R} and  \bibinfo{author}{Springel V}
      (\bibinfo{year}{2016}{\natexlab{a}}).
    \bibinfo{title}{Hydrodynamic moving-mesh simulations of the common envelope
      phase in binary stellar systems}.
    \bibinfo{journal}{{\em Astrophysical Journal, Letters}} \bibinfo{volume}{816}
      (\bibinfo{number}{1}): \bibinfo{pages}{L9}.
    
    \bibtype{Article}%
    \bibitem[Ohlmann et al.(2016{\natexlab{b}})]{ohlmann2016b}
    \bibinfo{author}{Ohlmann ST}, \bibinfo{author}{Röpke FK},
      \bibinfo{author}{Pakmor R}, \bibinfo{author}{Springel V} and
      \bibinfo{author}{Müller E} (\bibinfo{year}{2016}{\natexlab{b}}).
    \bibinfo{title}{Magnetic field amplification during the common envelope phase}.
    \bibinfo{journal}{{\em Monthly Notices of the Royal Astronomical Society}}
      \bibinfo{volume}{462} (\bibinfo{number}{1}): \bibinfo{pages}{L121--L125}.
    
    \bibtype{Article}%
    \bibitem[Ondratschek et al.(2022)]{ondratschek2022a}
    \bibinfo{author}{Ondratschek PA}, \bibinfo{author}{Röpke FK},
      \bibinfo{author}{Schneider FRN}, \bibinfo{author}{Fendt C},
      \bibinfo{author}{Sand C}, \bibinfo{author}{Ohlmann ST},
      \bibinfo{author}{Pakmor R} and  \bibinfo{author}{Springel V}
      (\bibinfo{year}{2022}), \bibinfo{month}{Apr.}
    \bibinfo{title}{Bipolar planetary nebulae from common-envelope evolution of
      binary stars}.
    \bibinfo{journal}{{\em Astronomy and Astrophysics}} \bibinfo{volume}{660}:
      \bibinfo{pages}{L8}.
    
    \bibtype{Article}%
    \bibitem[Ostriker(1999)]{ostriker1999a}
    \bibinfo{author}{Ostriker EC} (\bibinfo{year}{1999}), \bibinfo{month}{Mar.}
    \bibinfo{title}{Dynamical friction in a gaseous medium}.
    \bibinfo{journal}{{\em Astrophysical Journal}} \bibinfo{volume}{513}:
      \bibinfo{pages}{252--258}.
    
    \bibtype{Article}%
    \bibitem[Packet(1981)]{packet1981a}
    \bibinfo{author}{Packet W} (\bibinfo{year}{1981}), \bibinfo{month}{Sep.}
    \bibinfo{title}{On the spin-up of the mass accreting component in a close
      binary system}.
    \bibinfo{journal}{{\em Astronomy and Astrophysics}} \bibinfo{volume}{102}:
      \bibinfo{pages}{17--19}.
    
    \bibtype{Inproceedings}%
    \bibitem[Paczyński(1976)]{paczynski1976a}
    \bibinfo{author}{Paczyński B} (\bibinfo{year}{1976}), \bibinfo{month}{Jan.},
      \bibinfo{title}{Common envelope binaries}, \bibinfo{editor}{Eggleton P},
      \bibinfo{editor}{Mitton S} and  \bibinfo{editor}{Whelan J}, (Eds.),
      \bibinfo{booktitle}{Structure and evolution of close binary systems},
      \bibinfo{series}{{IAU} symposium}, \bibinfo{volume}{73},
      pp.~\bibinfo{pages}{75}.
    
    \bibtype{Article}%
    \bibitem[Pakmor et al.(2024)]{pakmor2024a}
    \bibinfo{author}{Pakmor R}, \bibinfo{author}{Pelisoli I},
      \bibinfo{author}{Justham S}, \bibinfo{author}{Rajamuthukumar AS},
      \bibinfo{author}{Röpke FK}, \bibinfo{author}{Schneider FRN},
      \bibinfo{author}{de~Mink SE}, \bibinfo{author}{Ohlmann ST},
      \bibinfo{author}{Podsiadlowski P}, \bibinfo{author}{Morán-Fraile J},
      \bibinfo{author}{Vetter M} and  \bibinfo{author}{Andrassy R}
      (\bibinfo{year}{2024}), \bibinfo{month}{Nov.}
    \bibinfo{title}{Large-scale ordered magnetic fields generated in mergers of
      helium white dwarfs}.
    \bibinfo{journal}{{\em Astronomy and Astrophysics}} \bibinfo{volume}{691}:
      \bibinfo{pages}{A179}.
    
    \bibtype{Article}%
    \bibitem[Passy et al.(2012)]{passy2012a}
    \bibinfo{author}{Passy JC}, \bibinfo{author}{De~Marco O},
      \bibinfo{author}{Fryer CL}, \bibinfo{author}{Herwig F},
      \bibinfo{author}{Diehl S}, \bibinfo{author}{Oishi JS},
      \bibinfo{author}{Mac~Low MM}, \bibinfo{author}{Bryan GL} and
      \bibinfo{author}{Rockefeller G} (\bibinfo{year}{2012}), \bibinfo{month}{Jan.}
    \bibinfo{title}{Simulating the common envelope phase of a red giant using
      smoothed-particle hydrodynamics and uniform-grid codes}.
    \bibinfo{journal}{{\em Astrophysical Journal}} \bibinfo{volume}{744}:
      \bibinfo{pages}{52}.
    
    \bibtype{Article}%
    \bibitem[Pastorello et al.(2019)]{pastorello2019b}
    \bibinfo{author}{Pastorello A}, \bibinfo{author}{Mason E},
      \bibinfo{author}{Taubenberger S}, \bibinfo{author}{Fraser M},
      \bibinfo{author}{Cortini G}, \bibinfo{author}{Tomasella L},
      \bibinfo{author}{Botticella MT}, \bibinfo{author}{Elias-Rosa N},
      \bibinfo{author}{Kotak R}, \bibinfo{author}{Smartt SJ},
      \bibinfo{author}{Benetti S}, \bibinfo{author}{Cappellaro E},
      \bibinfo{author}{Turatto M}, \bibinfo{author}{Tartaglia L},
      \bibinfo{author}{Djorgovski SG}, \bibinfo{author}{Drake AJ},
      \bibinfo{author}{Berton M}, \bibinfo{author}{Briganti F},
      \bibinfo{author}{Brimacombe J}, \bibinfo{author}{Bufano F},
      \bibinfo{author}{Cai YZ}, \bibinfo{author}{Chen S},
      \bibinfo{author}{Christensen EJ}, \bibinfo{author}{Ciabattari F},
      \bibinfo{author}{Congiu E}, \bibinfo{author}{Dimai A},
      \bibinfo{author}{Inserra C}, \bibinfo{author}{Kankare E},
      \bibinfo{author}{Magill L}, \bibinfo{author}{Maguire K},
      \bibinfo{author}{Martinelli F}, \bibinfo{author}{Morales-Garoffolo A},
      \bibinfo{author}{Ochner P}, \bibinfo{author}{Pignata G},
      \bibinfo{author}{Reguitti A}, \bibinfo{author}{Sollerman J},
      \bibinfo{author}{Spiro S}, \bibinfo{author}{Terreran G} and
      \bibinfo{author}{Wright DE} (\bibinfo{year}{2019}), \bibinfo{month}{Oct.}
    \bibinfo{title}{Luminous red novae: {Stellar} mergers or giant eruptions?}
    \bibinfo{journal}{{\em Astronomy and Astrophysics}} \bibinfo{volume}{630}:
      \bibinfo{pages}{A75}.
    
    \bibtype{Article}%
    \bibitem[Podsiadlowski and Joss(1989)]{podsiadlowski1989a}
    \bibinfo{author}{Podsiadlowski P} and  \bibinfo{author}{Joss PC}
      (\bibinfo{year}{1989}), \bibinfo{month}{Mar.}
    \bibinfo{title}{An alternative binary model for {SN1987A}}.
    \bibinfo{journal}{{\em Nature}} \bibinfo{volume}{338} (\bibinfo{number}{6214}):
      \bibinfo{pages}{401--403}.
    
    \bibtype{Article}%
    \bibitem[Podsiadlowski et al.(1990)]{podsiadlowski1990a}
    \bibinfo{author}{Podsiadlowski P}, \bibinfo{author}{Joss PC} and
      \bibinfo{author}{Rappaport S} (\bibinfo{year}{1990}), \bibinfo{month}{Jan.}
    \bibinfo{title}{A merger model for {SN} 1987 {A}}.
    \bibinfo{journal}{{\em Astronomy and Astrophysics}} \bibinfo{volume}{227}:
      \bibinfo{pages}{L9}.
    
    \bibtype{Article}%
    \bibitem[Podsiadlowski et al.(1992)]{podsiadlowski1992a}
    \bibinfo{author}{Podsiadlowski P}, \bibinfo{author}{Joss PC} and
      \bibinfo{author}{Hsu JJL} (\bibinfo{year}{1992}), \bibinfo{month}{May}.
    \bibinfo{title}{Presupernova evolution in massive interacting binaries}.
    \bibinfo{journal}{{\em Astrophysical Journal}} \bibinfo{volume}{391}:
      \bibinfo{pages}{246--264}.
    
    \bibtype{Article}%
    \bibitem[Prust and Chang(2019)]{prust2019a}
    \bibinfo{author}{Prust LJ} and  \bibinfo{author}{Chang P}
      (\bibinfo{year}{2019}), \bibinfo{month}{Jul.}
    \bibinfo{title}{Common envelope evolution on a moving mesh}.
    \bibinfo{journal}{{\em Monthly Notices of the Royal Astronomical Society}}
      \bibinfo{volume}{486} (\bibinfo{number}{4}): \bibinfo{pages}{5809--5818}.
    
    \bibtype{Article}%
    \bibitem[Renzo et al.(2020)]{renzo2020c}
    \bibinfo{author}{Renzo M}, \bibinfo{author}{Cantiello M},
      \bibinfo{author}{Metzger BD} and  \bibinfo{author}{Jiang YF}
      (\bibinfo{year}{2020}), \bibinfo{month}{Dec.}
    \bibinfo{title}{The stellar merger scenario for black holes in the
      pair-instability gap}.
    \bibinfo{journal}{{\em Astrophysical Journal, Letters}} \bibinfo{volume}{904}
      (\bibinfo{number}{2}): \bibinfo{pages}{L13}.
    
    \bibtype{Article}%
    \bibitem[Ricker and Taam(2012)]{ricker2012a}
    \bibinfo{author}{Ricker PM} and  \bibinfo{author}{Taam RE}
      (\bibinfo{year}{2012}), \bibinfo{month}{Feb.}
    \bibinfo{title}{An {AMR} study of the common-envelope phase of binary
      evolution}.
    \bibinfo{journal}{{\em Astrophysical Journal}} \bibinfo{volume}{746}:
      \bibinfo{pages}{74}.
    
    \bibtype{Article}%
    \bibitem[Rui and Fuller(2021)]{rui2021a}
    \bibinfo{author}{Rui NZ} and  \bibinfo{author}{Fuller J}
      (\bibinfo{year}{2021}), \bibinfo{month}{Dec.}
    \bibinfo{title}{Asteroseismic fingerprints of stellar mergers}.
    \bibinfo{journal}{{\em Monthly Notices of the Royal Astronomical Society}}
      \bibinfo{volume}{508} (\bibinfo{number}{2}): \bibinfo{pages}{1618--1631}.
    
    \bibtype{Article}%
    \bibitem[Röpke and De~Marco(2023)]{ropke2023a}
    \bibinfo{author}{Röpke FK} and  \bibinfo{author}{De~Marco O}
      (\bibinfo{year}{2023}), \bibinfo{month}{Dec.}
    \bibinfo{title}{Simulations of common-envelope evolution in binary stellar
      systems: physical models and numerical techniques}.
    \bibinfo{journal}{{\em Living Reviews in Computational Astrophysics}}
      \bibinfo{volume}{9} (\bibinfo{number}{1}): \bibinfo{pages}{2}.
    
    \bibtype{Article}%
    \bibitem[Sand et al.(2020)]{sand2020a}
    \bibinfo{author}{Sand C}, \bibinfo{author}{Ohlmann ST},
      \bibinfo{author}{Schneider FRN}, \bibinfo{author}{Pakmor R} and
      \bibinfo{author}{Röpke FK} (\bibinfo{year}{2020}), \bibinfo{month}{Dec.}
    \bibinfo{title}{Common-envelope evolution with an asymptotic giant branch
      star}.
    \bibinfo{journal}{{\em Astronomy and Astrophysics}} \bibinfo{volume}{644}:
      \bibinfo{pages}{A60}.
    
    \bibtype{Article}%
    \bibitem[Schneider et al.(2014)]{schneider2014a}
    \bibinfo{author}{Schneider FRN}, \bibinfo{author}{Izzard RG},
      \bibinfo{author}{de~Mink SE}, \bibinfo{author}{Langer N},
      \bibinfo{author}{Stolte A}, \bibinfo{author}{de~Koter A},
      \bibinfo{author}{Gvaramadze VV}, \bibinfo{author}{Hußmann B},
      \bibinfo{author}{Liermann A} and  \bibinfo{author}{Sana H}
      (\bibinfo{year}{2014}), \bibinfo{month}{Jan.}
    \bibinfo{title}{Ages of young star clusters, massive blue stragglers, and the
      upper mass limit of stars: {Analyzing} age-dependent stellar mass functions}.
    \bibinfo{journal}{{\em Astrophysical Journal}} \bibinfo{volume}{780}:
      \bibinfo{pages}{117}.
    
    \bibtype{Article}%
    \bibitem[Schneider et al.(2015)]{schneider2015a}
    \bibinfo{author}{Schneider FRN}, \bibinfo{author}{Izzard RG},
      \bibinfo{author}{Langer N} and  \bibinfo{author}{de~Mink SE}
      (\bibinfo{year}{2015}), \bibinfo{month}{May}.
    \bibinfo{title}{Evolution of mass functions of coeval stars through wind mass
      loss and binary interactions}.
    \bibinfo{journal}{{\em Astrophysical Journal}} \bibinfo{volume}{805}
      (\bibinfo{number}{1}): \bibinfo{pages}{20}.
    
    \bibtype{Article}%
    \bibitem[Schneider et al.(2016)]{schneider2016a}
    \bibinfo{author}{Schneider FRN}, \bibinfo{author}{Podsiadlowski P},
      \bibinfo{author}{Langer N}, \bibinfo{author}{Castro N} and
      \bibinfo{author}{Fossati L} (\bibinfo{year}{2016}), \bibinfo{month}{Apr.}
    \bibinfo{title}{Rejuvenation of stellar mergers and the origin of magnetic
      fields in massive stars}.
    \bibinfo{journal}{{\em Monthly Notices of the Royal Astronomical Society}}
      \bibinfo{volume}{457}: \bibinfo{pages}{2355--2365}.
    
    \bibtype{Article}%
    \bibitem[Schneider et al.(2019)]{schneider2019a}
    \bibinfo{author}{Schneider FRN}, \bibinfo{author}{Ohlmann ST},
      \bibinfo{author}{Podsiadlowski P}, \bibinfo{author}{Röpke FK},
      \bibinfo{author}{Balbus SA}, \bibinfo{author}{Pakmor R} and
      \bibinfo{author}{Springel V} (\bibinfo{year}{2019}), \bibinfo{month}{Oct.}
    \bibinfo{title}{Stellar mergers as the origin of magnetic massive stars}.
    \bibinfo{journal}{{\em Nature}} \bibinfo{volume}{574} (\bibinfo{number}{7777}):
      \bibinfo{pages}{211--214}.
    
    \bibtype{Article}%
    \bibitem[Schneider et al.(2020)]{schneider2020a}
    \bibinfo{author}{Schneider FRN}, \bibinfo{author}{Ohlmann ST},
      \bibinfo{author}{Podsiadlowski P}, \bibinfo{author}{Röpke FK},
      \bibinfo{author}{Balbus SA} and  \bibinfo{author}{Pakmor R}
      (\bibinfo{year}{2020}), \bibinfo{month}{May}.
    \bibinfo{title}{Long-term evolution of a magnetic massive merger product}.
    \bibinfo{journal}{{\em Monthly Notices of the Royal Astronomical Society}}
      \bibinfo{volume}{495} (\bibinfo{number}{3}): \bibinfo{pages}{2796--2812}.
    
    \bibtype{Article}%
    \bibitem[Schneider et al.(2024)]{schneider2024a}
    \bibinfo{author}{Schneider FRN}, \bibinfo{author}{Podsiadlowski P} and
      \bibinfo{author}{Laplace E} (\bibinfo{year}{2024}), \bibinfo{month}{Jun.}
    \bibinfo{title}{Pre-supernova evolution and final fate of stellar mergers and
      accretors of binary mass transfer}.
    \bibinfo{journal}{{\em Astronomy and Astrophysics}} \bibinfo{volume}{686}:
      \bibinfo{pages}{A45}.
    
    \bibtype{Article}%
    \bibitem[Schwab et al.(2012)]{schwab2012a}
    \bibinfo{author}{Schwab J}, \bibinfo{author}{Shen KJ},
      \bibinfo{author}{Quataert E}, \bibinfo{author}{Dan M} and
      \bibinfo{author}{Rosswog S} (\bibinfo{year}{2012}), \bibinfo{month}{Nov.}
    \bibinfo{title}{The viscous evolution of white dwarf merger remnants}.
    \bibinfo{journal}{{\em Monthly Notices of the Royal Astronomical Society}}
      \bibinfo{volume}{427} (\bibinfo{number}{1}): \bibinfo{pages}{190--203}.
    
    \bibtype{Article}%
    \bibitem[Shiber et al.(2017)]{shiber2017a}
    \bibinfo{author}{Shiber S}, \bibinfo{author}{Kashi A} and
      \bibinfo{author}{Soker N} (\bibinfo{year}{2017}), \bibinfo{month}{Feb.}
    \bibinfo{title}{Simulating the onset of grazing envelope evolution of binary
      stars}.
    \bibinfo{journal}{{\em Monthly Notices of the Royal Astronomical Society}}
      \bibinfo{volume}{465} (\bibinfo{number}{1}): \bibinfo{pages}{L54--L58}.
    
    \bibtype{Article}%
    \bibitem[Shiber et al.(2019)]{shiber2019a}
    \bibinfo{author}{Shiber S}, \bibinfo{author}{Iaconi R},
      \bibinfo{author}{De~Marco O} and  \bibinfo{author}{Soker N}
      (\bibinfo{year}{2019}), \bibinfo{month}{Oct.}
    \bibinfo{title}{Companion-launched jets and their effect on the dynamics of
      common envelope interaction simulations}.
    \bibinfo{journal}{{\em Monthly Notices of the Royal Astronomical Society}}
      \bibinfo{volume}{488} (\bibinfo{number}{4}): \bibinfo{pages}{5615--5632}.
    
    \bibtype{Article}%
    \bibitem[Sills et al.(1997)]{sills1997a}
    \bibinfo{author}{Sills A}, \bibinfo{author}{Lombardi Jr. JC},
      \bibinfo{author}{Bailyn CD}, \bibinfo{author}{Demarque P},
      \bibinfo{author}{Rasio FA} and  \bibinfo{author}{Shapiro SL}
      (\bibinfo{year}{1997}), \bibinfo{month}{Sep.}
    \bibinfo{title}{Evolution of stellar collision products in globular clusters.
      {I}. {Head}-on collisions}.
    \bibinfo{journal}{{\em Astrophysical Journal}} \bibinfo{volume}{487}:
      \bibinfo{pages}{290}.
    
    \bibtype{Article}%
    \bibitem[Smith et al.(2018)]{smith2018a}
    \bibinfo{author}{Smith N}, \bibinfo{author}{Andrews JE}, \bibinfo{author}{Rest
      A}, \bibinfo{author}{Bianco FB}, \bibinfo{author}{Prieto JL},
      \bibinfo{author}{Matheson T}, \bibinfo{author}{James DJ},
      \bibinfo{author}{Smith RC}, \bibinfo{author}{Strampelli GM} and
      \bibinfo{author}{Zenteno A} (\bibinfo{year}{2018}), \bibinfo{month}{Oct.}
    \bibinfo{title}{Light echoes from the plateau in {Eta} {Carinae}'s {Great}
      {Eruption} reveal a two-stage shock-powered event}.
    \bibinfo{journal}{{\em Monthly Notices of the Royal Astronomical Society}}
      \bibinfo{volume}{480}: \bibinfo{pages}{1466--1498}.
    
    \bibtype{Article}%
    \bibitem[Soker(2015)]{soker2015a}
    \bibinfo{author}{Soker N} (\bibinfo{year}{2015}), \bibinfo{month}{Feb.}
    \bibinfo{title}{Close stellar binary systems by grazing envelope evolution}.
    \bibinfo{journal}{{\em Astrophysical Journal}} \bibinfo{volume}{800}
      (\bibinfo{number}{2}): \bibinfo{pages}{114}.
    
    \bibtype{Article}%
    \bibitem[Spitzer and Saslaw(1966)]{spitzer1966a}
    \bibinfo{author}{Spitzer Lyman J} and  \bibinfo{author}{Saslaw WC}
      (\bibinfo{year}{1966}), \bibinfo{month}{Feb.}
    \bibinfo{title}{On the evolution of galactic nuclei}.
    \bibinfo{journal}{{\em Astrophysical Journal}} \bibinfo{volume}{143}:
      \bibinfo{pages}{400}.
    
    \bibtype{Article}%
    \bibitem[Staff et al.(2016)]{staff2016a}
    \bibinfo{author}{Staff JE}, \bibinfo{author}{De~Marco O}, \bibinfo{author}{Wood
      P}, \bibinfo{author}{Galaviz P} and  \bibinfo{author}{Passy JC}
      (\bibinfo{year}{2016}), \bibinfo{month}{May}.
    \bibinfo{title}{Hydrodynamic simulations of the interaction between giant stars
      and planets}.
    \bibinfo{journal}{{\em Monthly Notices of the Royal Astronomical Society}}
      \bibinfo{volume}{458}: \bibinfo{pages}{832--844}.
    
    \bibtype{Article}%
    \bibitem[Taam(1979)]{taam1979a}
    \bibinfo{author}{Taam RE} (\bibinfo{year}{1979}), \bibinfo{month}{Jan.}
    \bibinfo{title}{Double core evolution and {X}-ray binaries}.
    \bibinfo{journal}{{\em Astrophysics Letters}}
    
    \bibtype{Article}%
    \bibitem[Taam et al.(1978)]{taam1978a}
    \bibinfo{author}{Taam RE}, \bibinfo{author}{Bodenheimer P} and
      \bibinfo{author}{Ostriker JP} (\bibinfo{year}{1978}), \bibinfo{month}{May}.
    \bibinfo{title}{Double core evolution. {I} - {A} 16 solar mass star with a 1
      solar mass neutron-star companion}.
    \bibinfo{journal}{{\em Astrophysical Journal}} \bibinfo{volume}{222}:
      \bibinfo{pages}{269--280}.
    
    \bibtype{Article}%
    \bibitem[Tylenda et al.(2011)]{tylenda2011a}
    \bibinfo{author}{Tylenda R}, \bibinfo{author}{Hajduk M},
      \bibinfo{author}{Kamiński T}, \bibinfo{author}{Udalski A},
      \bibinfo{author}{Soszyński I}, \bibinfo{author}{Szymański MK},
      \bibinfo{author}{Kubiak M}, \bibinfo{author}{Pietrzyński G},
      \bibinfo{author}{Poleski R}, \bibinfo{author}{Wyrzykowski L.} and
      \bibinfo{author}{Ulaczyk K} (\bibinfo{year}{2011}), \bibinfo{month}{Apr.}
    \bibinfo{title}{V1309 {Scorpii}: merger of a contact binary}.
    \bibinfo{journal}{{\em Astronomy and Astrophysics}} \bibinfo{volume}{528}:
      \bibinfo{pages}{A114}.
    
    \bibtype{Inproceedings}%
    \bibitem[van~den Heuvel(1976)]{vandenheuvel1976a}
    \bibinfo{author}{van~den Heuvel EPJ} (\bibinfo{year}{1976}),
      \bibinfo{month}{Jan.}, \bibinfo{title}{Late stages of close binary systems},
      \bibinfo{editor}{Eggleton P}, \bibinfo{editor}{Mitton S} and
      \bibinfo{editor}{Whelan J}, (Eds.), \bibinfo{booktitle}{Structure and
      evolution of close binary systems}, \bibinfo{volume}{73},
      pp.~\bibinfo{pages}{35}.
    
    \bibtype{Article}%
    \bibitem[Vanbeveren et al.(2013)]{vanbeveren2013a}
    \bibinfo{author}{Vanbeveren D}, \bibinfo{author}{Mennekens N},
      \bibinfo{author}{Van~Rensbergen W} and  \bibinfo{author}{De~Loore C}
      (\bibinfo{year}{2013}), \bibinfo{month}{Apr.}
    \bibinfo{title}{Blue supergiant progenitor models of type {II} supernovae}.
    \bibinfo{journal}{{\em Astronomy and Astrophysics}} \bibinfo{volume}{552}:
      \bibinfo{pages}{A105}.
    
    \bibtype{Article}%
    \bibitem[Vetter et al.(2024)]{vetter2024a}
    \bibinfo{author}{Vetter M}, \bibinfo{author}{Röpke FK},
      \bibinfo{author}{Schneider FRN}, \bibinfo{author}{Pakmor R},
      \bibinfo{author}{Ohlmann ST}, \bibinfo{author}{Lau MYM} and
      \bibinfo{author}{Andrassy R} (\bibinfo{year}{2024}), \bibinfo{month}{Nov.}
    \bibinfo{title}{From spherical stars to disk-like structures: {3D}
      common-envelope evolution of massive binaries beyond inspiral}.
    \bibinfo{journal}{{\em Astronomy and Astrophysics}} \bibinfo{volume}{691}:
      \bibinfo{pages}{A244}.
    
    \bibtype{Article}%
    \bibitem[von Zeipel(1910)]{vonzeipel1910a}
    \bibinfo{author}{von Zeipel H} (\bibinfo{year}{1910}), \bibinfo{month}{Mar.}
    \bibinfo{title}{Sur l'application des séries de {M}. {Lindstedt} à l'étude
      du mouvement des comètes périodiques}.
    \bibinfo{journal}{{\em Astronomische Nachrichten}} \bibinfo{volume}{183}:
      \bibinfo{pages}{345}.
    
    \bibtype{Article}%
    \bibitem[Wagg et al.(2024)]{wagg2024a}
    \bibinfo{author}{Wagg T}, \bibinfo{author}{Johnston C},
      \bibinfo{author}{Bellinger EP}, \bibinfo{author}{Renzo M},
      \bibinfo{author}{Townsend R} and  \bibinfo{author}{de~Mink SE}
      (\bibinfo{year}{2024}), \bibinfo{month}{Jul.}
    \bibinfo{title}{The asteroseismic imprints of mass transfer. {A} case study of
      a binary mass-gainer in the {SPB} instability strip}.
    \bibinfo{journal}{{\em Astronomy and Astrophysics}} \bibinfo{volume}{687}:
      \bibinfo{pages}{A222}.
    
    \bibtype{Article}%
    \bibitem[Wang et al.(2022)]{wang2022a}
    \bibinfo{author}{Wang C}, \bibinfo{author}{Langer N},
      \bibinfo{author}{Schootemeijer A}, \bibinfo{author}{Milone A},
      \bibinfo{author}{Hastings B}, \bibinfo{author}{Xu XT},
      \bibinfo{author}{Bodensteiner J}, \bibinfo{author}{Sana H},
      \bibinfo{author}{Castro N}, \bibinfo{author}{Lennon DJ},
      \bibinfo{author}{Marchant P}, \bibinfo{author}{de~Koter A} and
      \bibinfo{author}{de~Mink SE} (\bibinfo{year}{2022}), \bibinfo{month}{Feb.}
    \bibinfo{title}{Stellar mergers as the origin of the blue main-sequence band in
      young star clusters}.
    \bibinfo{journal}{{\em Nature Astronomy}} \bibinfo{volume}{6}:
      \bibinfo{pages}{480--487}.
    
    \bibtype{Phdthesis}%
    \bibitem[Webbink(1975)]{webbink1975a}
    \bibinfo{author}{Webbink RF} (\bibinfo{year}{1975}), \bibinfo{month}{Jul.}
    \bibinfo{title}{The evolution of low-mass close binary systems}.
    \bibinfo{journal}{Ph.D.\ Thesis}
    
    
    \bibtype{Article}%
    \bibitem[Webbink(1984)]{webbink1984a}
    \bibinfo{author}{Webbink RF} (\bibinfo{year}{1984}), \bibinfo{month}{Feb.}
    \bibinfo{title}{Double white dwarfs as progenitors of {R} {Coronae} {Borealis}
      stars and {Type} {I} supernovae}.
    \bibinfo{journal}{{\em Astrophysical Journal}} \bibinfo{volume}{277}:
      \bibinfo{pages}{355--360}.
    
    \bibtype{Article}%
    \bibitem[Wei et al.(2024)]{wei2024a}
    \bibinfo{author}{Wei D}, \bibinfo{author}{Schneider FRN},
      \bibinfo{author}{Podsiadlowski P}, \bibinfo{author}{Laplace E},
      \bibinfo{author}{Röpke FK} and  \bibinfo{author}{Vetter M}
      (\bibinfo{year}{2024}), \bibinfo{month}{Aug.}
    \bibinfo{title}{Evolution and final fate of massive post-common-envelope
      binaries}.
    \bibinfo{journal}{{\em Astronomy and Astrophysics}} \bibinfo{volume}{688}:
      \bibinfo{pages}{A87}.
    
    \bibtype{Article}%
    \bibitem[Wickramasinghe et al.(2014)]{wickramasinghe2014a}
    \bibinfo{author}{Wickramasinghe DT}, \bibinfo{author}{Tout CA} and
      \bibinfo{author}{Ferrario L} (\bibinfo{year}{2014}), \bibinfo{month}{Jan.}
    \bibinfo{title}{The most magnetic stars}.
    \bibinfo{journal}{{\em Monthly Notices of the Royal Astronomical Society}}
      \bibinfo{volume}{437}: \bibinfo{pages}{675--681}.
    
    \bibtype{Article}%
    \bibitem[Zorotovic et al.(2010)]{zorotovic2010a}
    \bibinfo{author}{Zorotovic M}, \bibinfo{author}{Schreiber MR},
      \bibinfo{author}{Gänsicke BT} and  \bibinfo{author}{Nebot Gómez-Morán A}
      (\bibinfo{year}{2010}), \bibinfo{month}{Sep.}
    \bibinfo{title}{Post-common-envelope binaries from {SDSS}. {IX}: {Constraining}
      the common-envelope efficiency}.
    \bibinfo{journal}{{\em Astronomy and Astrophysics}} \bibinfo{volume}{520}:
      \bibinfo{pages}{A86}.
    
    \end{thebibliography*}

\end{document}